\documentclass[10pt,a4paper,journal]{IEEEtran}
\usepackage{setspace}
\usepackage{epsfig}
\usepackage{times}
\usepackage{latexsym}
\usepackage{subfigure}
\usepackage{amsfonts,amsmath,amsthm,amssymb}
\usepackage{graphicx}
\usepackage{color}

\theoremstyle{definition}
\newtheorem{definition}{Definition}[section]

\theoremstyle{remark}
\newtheorem{remark}{Remarks}[section] %\label{rmk:}
\newtheorem{example}{Example}[section]

\theoremstyle{plain}
\newtheorem{theorem}{Theorem}[section]

\newcommand{\bmath}[1]{\mbox{\boldmath$#1$}}
\newcommand{\EXP}[1]{\mathsf{E}\!\left(#1\right)}
\newcommand{\ind}[1]{\mathbf{1}_{\left\{#1\right\}}}

\begin{document}

\title{Cell-Level Modeling of IEEE 802.11 WLANs.}

\author{\IEEEauthorblockN{Manoj Panda and Anurag Kumar} \\
\IEEEauthorblockA{ECE Department, IISc Bangalore -- 560012. \\ Email:
  \{manoj,anurag\}@ece.iisc.ernet.in}\thanks{This paper is an extended
  version of our earlier work
  \cite{wanet.manoj-anuragRAWNET09multicell}. In this paper, we extend
  our analytical model in \cite{wanet.manoj-anuragRAWNET09multicell}
  to TCP-controlled short-file downloads, characterize the associated
  service process, and provide approximations for the mean flow-transfer
  delays. We also demonstrate the applicability of our analytical model 
  when the ``PBD condition'' does not strictly hold.}}

\maketitle

\begin{abstract}
We develop a scalable \textit{cell-level} analytical model for multi-cell infrastructure IEEE 802.11 WLANs under a so-called Pairwise Binary Dependence (PBD) condition. The PBD condition is a geometric property under which the relative locations of the nodes inside a cell do not matter and the network is free of \textit{hidden nodes}. For the cases of saturated nodes and TCP-controlled long-file downloads, we provide accurate predictions of cell throughputs. Similar to Bonald et al (Sigmetrics, 2008), we model a multi-cell WLAN under short-file downloads as ``a network of processor-sharing queues with state-dependent service rates.'' Whereas the state-dependent service rates proposed by Bonald et al are based only on the \textit{number} of contending neighbors, we employ state-dependent service rates that incorporate the the impact of the overall \textit{topology} of the network. We propose an \textit{effective service rate approximation} technique and obtain good approximations for the \textit{mean flow transfer delay} in each cell. For TCP-controlled downloads where the APs transmit a large fraction of time, we show that the throughputs predicted under the PBD condition are very good approximations in two important scenarios where hidden nodes are indeed present and the PBD condition does not strictly hold. 

\keywords{IEEE 802.11 \and multi-cell WLAN \and analytical modeling \and throughput and delay}
\end{abstract}

\section{Introduction}
\label{sec:introduction}

With widespread deployment of WiFi networks (or, more formally, IEEE 802.11 networks) in office buildings, university campuses, homes, hotels, airports and other public places, it has become very important to understand the performance of Wireless Local Area Networks (WLANs) that are based on the IEEE 802.11 standard, and also to know how to effectively design, deploy and manage them. 

The IEEE 802.11 standard \cite{wanet.IEEE802dot11standard2007} provides two modes of operation, namely, the \textit{ad hoc} mode and the \textit{infrastructure} mode. Commercial and enterprise WLANs usually operate in the infrastructure mode. An infrastructure WLAN contains one or more Access Points (APs) which provide service to a set of users or client stations (STAs). Every STA in the WLAN associates iself with exactly one AP. Each AP, along with its associated STAs, constitutes a so-called \textit{cell}. Each cell operates on a specific channel. Cells that operate on the same channel are called \textit{co-channel}. We call a WLAN containing multiple APs a \textit{multi-cell} WLAN. 

\begin{figure}[tb]
\centering \
\begin{minipage}{7.8cm}
\begin{center}
\includegraphics[scale=0.25]{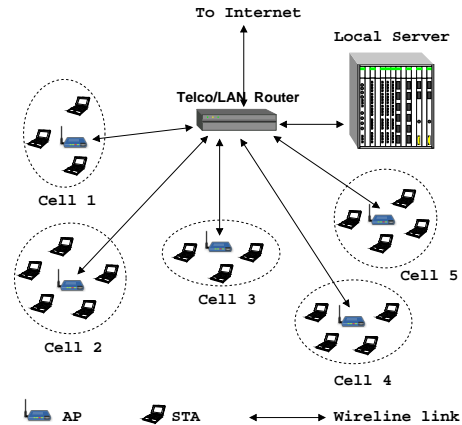}
\caption{A multi-cell infrastructure WLAN: The 802.11 MAC protocol is employed only for single-hop intra-cell frame exchanges within the cells (shown by dashed ovals). A high-speed wireline LAN connects the APs and provides access to the Internet. \label{fig:multicell-infrastructure-wlan}}
\end{center}
\end{minipage}
%\vspace{-5mm}
\end{figure}

In a multi-cell infrastructure WLAN, the APs are usually connected among themselves and to the Internet by a high-speed \textit{wireline} Local Area Network (LAN), e.g., a Gigabit Ethernet. The STAs, on the other hand, access the Internet \textit{only through their respective APs}. Thus, the 802.11 MAC protocol is employed only for \textit{single-hop intra-cell} frame exchanges between an AP and its associated STAs (and not among the APs and STAs belonging to different cells). Figure \ref{fig:multicell-infrastructure-wlan} depicts an example of such a multi-cell WLAN.

This paper is concerned with analytical modeling of infrastructure WLANs (such as in Figure \ref{fig:multicell-infrastructure-wlan}) that are based on the Distributed Coordination Function (DCF) Medium Access Control (MAC) protocol as defined in the IEEE 802.11 standard \cite{wanet.IEEE802dot11standard2007}. Analytical modeling can provide important insights to effectively design, deploy and manage the WLANs. However, accurate analytical modeling of multi-cell WLANs is a challenging problem. Nodes (i.e.~AP or STA) in two closely located co-channel cells can suppress each other's transmissions via carrier sensing and interfere with each other's receptions causing packet losses. Thus, the activities of nodes in proximal co-channel cells are essentially coupled, which makes the analytical modeling difficult.

\begin{figure}[tb]
\centering \
\begin{minipage}{7.8cm}
\begin{center}
\includegraphics[scale=0.28]{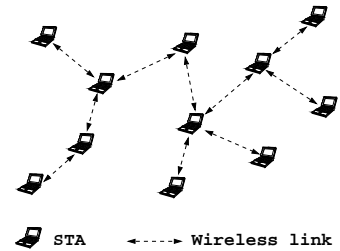}
\caption{A multi-hop ad hoc WLAN. \label{fig:multihopAdHoc}}
\end{center}
\end{minipage}
%\vspace{-5mm}
%The 802.11 MAC protocol is employed only for single-hop intra-cell frame exchanges within the cells (shown by dashed ovals). 
\end{figure}

\subsection{Literature Survey}
\label{subsec:literature-survey}

%Nodes in a multi-cell infrastructure WLAN, however, can sense and/or decode the transmissions of different subsets of APs and other STAs \cite{wanet.bonald08multicellprocsharing}. Thus, n

The seminal analytical model for \textit{single cell} WLANs was developed by \cite{wanet.bianchi00performance}, and later generalized by \cite{wanet.kumar_etal07new_insights}. In a single cell, nodes can sense and decode each other's transmissions. Thus, nodes in a single cell have the same \textit{global} view of the activities on the common medium. Nodes in a multi-cell WLAN, however, can have different \textit{local} views of the network activity around themselves, and their own activity is determined by this local view. 

In the context of multi-hop ad hoc networks (see Figure \ref{fig:multihopAdHoc}), \textit{node-} and \textit{link-level} models have been proposed to capture the \textit{local} characteristics of network activities \cite{wanet.boorstyn87multihop,wanet.wang-kar05multihop,wanet.garetto_etal08starvation}. However, for networks of realistic size, a node- or link-level model is intractable, since its complexity increases exponentially with the number of nodes or links \cite{wanet.kershenbaum-etal87complex}. 

Since a node- or link-level model is often intractable and provides little insight, several simplifying assumptions have been made in order to gain insights. A common simplification is to ignore collisions due to \textit{hidden nodes} by assuming that hidden nodes are suppressed either via RTS/CTS handshaking \cite{wanet.durvy09selfOrganization,wanet.bonald08multicellprocsharing}, or via a so-called ``hidden-node-free design'' \cite{wanet.jiang-liew08MobComp-HNEN,wanet.liew_etal09mobicom-capacity-wireless-networks,wanet.liew_etal09ICCback-of-the-envelope}. Another simplification is to assume an infinite number of nodes placed according to either some regular topology \cite{wanet.durvy09selfOrganization} or a regular point process \cite{wanet.nguyen07stochGeom}. 

TCP-controlled data transfers are usually classified into two types: (i) \textit{long-lived} flows (e.g., file transfer), and (ii) \textit{short-lived} flows (e.g., web browsing). The case of long-lived flows in a single isolated cell has been analyzed in \cite{wanet.harsha07WiNet,wanet.bruno08TCPeqvSatModel}. A single isolated cell with short-lived flows was modeled as ``a processor-sharing queue with state-dependent service rates'' in \cite{wanet.litjens_etalITC03integrated_packet_flow,wanet.miorandi_etal06http_over_wlans}. The seminal papers by Bonald et al.~provide the most comprehensive treatment of multi-cell networks with short-lived flows, both for cellular data networks \cite{wanet.bonald_etal04multicell-cellular} as well as for 802.11-based WLANs \cite{wanet.bonald08multicellprocsharing}. \cite{wanet.bonald08multicellprocsharing} model a multi-cell infrastructure WLAN under short-lived downloads as a ``network of \textit{multi-class} processor-sharing queues with state-dependent service rates.''

\subsection{Our Contributions} 

To accurately model a multi-cell infrastructure WLAN, one needs to account for the node- and link-level interactions, as in the case of multi-hop ad hoc networks. However, comparing Figures \ref{fig:multicell-infrastructure-wlan} and \ref{fig:multihopAdHoc}, we observe a key difference: 
\begin{itemize}
\item The ad hoc WLAN does not possess any specific stucture (topology). However, the infrastructure WLAN has a nice \textit{cellular} structure. For communication at high Physical layer (PHY) rate to be possible, the \textit{cell radius} $R$ (i.e., the maximum distance between an AP and its associated STAs) would be small. However, with multiple non-overlapping channels, the distance $D$ between two nearest co-channel APs would be larger than $2R$. Thus, the set of co-channel cells (corresponding to some specific channel) will have a cellular structure (for example, see Figures \ref{fig:twoCell}-\ref{fig:sevenCellArbitrary}). 
\end{itemize}

The above observation motivates us to take a different perspective on the analytical modeling of multi-cell infrastructure WLANs. Specifically, we exploit the cellular structure of multi-cell infrastructure WLANs by treating a cell as a single entity and develop a \textit{cell-level} analytical model. We make the following novel contributions: 

\vspace{1mm}

\noindent \textbf{(1)} We identify a geometric property, which we call the \textit{Pairwise Binary Dependence} (PBD) condition (see A1 in Section \ref{sec:model-assumptions}), under which we develop a \textit{cell-level} model. Our cell-level model is \textit{scalable} since its complexity increases with number of cells rather than nodes or links (see Remark \ref{rmk:model-complexity}). 

\vspace{1mm}

\noindent \textbf{(2)} We develop analytical models with saturated AP and STA queues as well as for TCP-controlled long- and short-file downloads (Sections \ref{sec:analysis-multicell-arbitrary-cell-topology} and \ref{sec:short-lived}). For the case of saturated nodes and TCP-controlled long-lived downloads (resp. short-lived downloads), we predict throughputs (resp.~flow transfer delays) and illustrate the accuracy of our analysis (Section \ref{sec:results}). The expressions of collision probabilities, throughputs and mean flow transfer delays that we derive are new. 

\vspace{1mm}

\noindent \textbf{(3)} The PBD condition essentially makes a multi-cell WLAN free of hidden nodes (Section \ref{sec:model-assumptions}). We develop our analytical model in a hidden-node-free setting as in \cite{wanet.durvy09selfOrganization,wanet.bonald08multicellprocsharing,wanet.jiang-liew08MobComp-HNEN,wanet.liew_etal09mobicom-capacity-wireless-networks,wanet.liew_etal09ICCback-of-the-envelope}. However, we also demonstrate the applicability of our analytical model when hidden nodes are indeed present and the PBD condition does not strictly hold (Section \ref{subsec:when-PBD-does-not-hold}). Specifically, we show that the throughput predictions provided by our analytical model are reasonably accurate also when: 

\vspace{1mm}

\noindent (i) The AP in a cell can sense and decode the transmissions of all the STAs in a neighboring cell, but $\approx 50\%$ of the STAs in a cell can sense and decode the transmissions of only $\approx 50\%$ of the STAs in the neighboring cell, and 

\vspace{1mm}

\noindent (ii) The AP in a cell can sense and decode the transmissions of the AP and only $\approx 50\%$ of the STAs in the other cell. 

\vspace{1mm}

\noindent \textbf{(4)} In \cite{wanet.wang-kar05multihop,wanet.durvy09selfOrganization,wanet.liew_etal09ICCback-of-the-envelope}, valuable insights have been obtained by taking the \textit{access intensity}\footnote{This term was coined by \cite{wanet.durvy09selfOrganization} and is defined in \eqref{eqn:varrho-definition}.} to infinity. One such insight is:
\begin{itemize}
\item [I$_1$] As the access intensity goes to infinity, only those nodes that belong to some \textit{maximum independent set}\footnote{A maximum independent set of a graph $G$ is an independent set of $G$ with maximum cardinality. In contrast, a \textit{maximal} independent set of $G$ is an independent set of $G$ that is not a proper subset of any other independent set of $G$. Every maximum set is also a maximal set, but the converse is not true.} obtain non-zero throughputs \cite{wanet.wang-kar05multihop,wanet.durvy09selfOrganization,wanet.liew_etal09ICCback-of-the-envelope}.
\end{itemize}
We confirm Insight (I$_1$) at the cell-level. Another insight, which is a simple consequence of Insight (I$_1$), is: 
\begin{itemize}
\item [I$_2$] As the access intensity goes to infinity, the aggregate network throughput is maximized.
\end{itemize} 
Insight (I$_2$) has been recently established by \cite{wanet.durvy09selfOrganization} in the context of an infinite linear chain of saturated nodes. We establish it, under the PBD condition, for arbitrary cell topologies for saturated nodes and also for TCP-controlled long-lived downloads (see Theorem \ref{thm:maximum-throughput}). Moreover, we provide a third insight:
\begin{itemize}
\item [I$_3$] Realistic access intensities with default backoff parameter settings and sufficiently large packet payloads (e.g., $\geq$ 500 bytes) are indeed \textit{large} so that the values of the cell throughputs in such situations are very close to their values with infinite access intensities (Section \ref{subsec:large-rho-regime}).
\end{itemize} 

\vspace{1mm}

\noindent \textbf{(5)} For TCP-controlled short-file downloads, we improve the service model of \cite{wanet.bonald08multicellprocsharing}. The service model in \cite{wanet.bonald08multicellprocsharing} is based on the assumption of \textit{``synchronous and slotted''} evolution of network activities with the service rates depending only on the \textit{number} of contending neighbors. However, inter-cell blocking due to carrier sensing is ``asynchronous'' (see Section \ref{subsec:saturated-case}). The service rates in our model capture this asynchrony and the service rates in our model depend on the overall \textit{topology} of the contending APs. Taking an example of a three-cell network, we show that (i) the service rates proposed in \cite{wanet.bonald08multicellprocsharing} can lead to inaccurate prediction of delays (Figure \ref{fig:shortTCPDelayVsEVNuPoint1ThreeCellLinearModel1}), but (ii) significantly more accurate results can be obtained with our improved service model (Figure \ref{fig:shortTCPDelayVsEVNuPoint1ThreeCellLinearModel2}). 

%\vspace{2mm}

%\noindent \textbf{(5)} We provide an approximate delay analysis to predict the mean flow transfer delays in each cell (Section \ref{subsubsec:delay-analysis}). 

\subsection{Outline of the paper} 

The remainder of the paper is structured as follows. In Section \ref{sec:model-assumptions}, we provide our network model and discuss our assumptions. In Section \ref{sec:analysis-multicell-arbitrary-cell-topology}, we first develop our analytical model with saturated AP and STA queues, and then extend to the case of TCP-controlled long-file downloads (Section \ref{subsec:TCP-traffic}). In Section \ref{sec:short-lived}, we apply the insights obtained in Section \ref{subsec:large-rho-regime} to extend to the case of TCP-controlled short-file downloads. We illustrate the accuracy of our analytical models in Section \ref{sec:results}. The paper concludes in Section \ref{sec:conclusion}. A derivation of the equation for collision probability has been provided in the Appendix at the end of the paper.

\section{Network Model and Assumptions}
\label{sec:model-assumptions}

%Let $R$ denote the \textit{cell radius}, i.e., $R$ is the maximum distance between an AP and the STAs associated with it. 

Let $R_{cs}$ denote the \textit{carrier sensing range}, i.e., $R_{cs}$ is the maximum distance up to which a transmitter can cause a ``medium busy'' condition at idle receivers \cite{wanet.roy_etal09ToN-PCS,wanet.liew_etal09mobicom-capacity-wireless-networks}. 

\begin{definition}
\label{defn:dependence}
Two nodes are said to be \textit{dependent} if the distance between them is less than $R_{cs}$ \textbf{and} they operate on the same channel; otherwise, the two nodes are said to be \textit{independent}. Two cells are said to be independent if every node in a cell is independent w.r.t.~every node in the other cell; otherwise, the two cells are said to be dependent. Two dependent cells are said to be \textit{completely dependent} if every node in a cell is dependent w.r.t.~every node in the other cell. \hfill \qed
\end{definition}

In the above definitions, we implicitly assume that (i) only non-overlapping channels are used, (ii) nodes can neither sense nor interfere (and be interfered) with transmissions on a different non-overlapping channel, and (iii) the carrier sensing range $R_{cs}$ is a \textit{sharp} boundary within which nodes sense and interfere (and be interfered) with each other's transmissions, and do not interact outside of it. 

We make the following assumptions about the network:
\begin{itemize}

\item [A0] Nodes in the same cell are dependent. 

\item [A1] \textbf{Pairwise Binary Dependence (PBD):} Any pair of cells is either independent or completely dependent. 

\item [A2] Simultaneous transmissions by independent transmitters are successfully received at their respective receivers. 

\item [A3] Simultaneous transmissions by dependent transmitters are always lost due to collision. 

\item [A4] Each cell contains a \textit{fixed} number of \textit{identical} STAs. 

\item [A5] Packet losses due to channel errors are negligible.

\end{itemize}

%\subsection{Implications and Justifications}

By A0, nodes in the same cell can sense and interfere (and be interfered) with each other's transmission. By the PBD condition (A1), nodes in the same cell sense and interfere (and be interfered) with the same set of nodes in the other cells. Since communication is always between an AP and one of its associated STAs (in the same cell), any transmitter-receiver pair can sense the same set of interferers. Thus, the PBD condition implies that the network is free of \textit{hidden nodes}. As stated earlier in Section \ref{subsec:literature-survey}, assuming a hidden-node-free setting is a common simplification and also applied in \cite{wanet.durvy09selfOrganization,wanet.bonald08multicellprocsharing,wanet.jiang-liew08MobComp-HNEN,wanet.liew_etal09mobicom-capacity-wireless-networks,wanet.liew_etal09ICCback-of-the-envelope}. In Section \ref{subsec:when-PBD-does-not-hold} we show that our cell-level model, which we develop assuming the PBD condition, can also provide very good approximations in certain scenarios in which hidden nodes are indeed present and the PBD condition does not strictly hold. 

Simultaneous transmissions by nodes that are located within $R_{cs}$ of each other (i.e., nodes that sense each other) lead to \textit{synchronous} collisions, in which case, the colliding transmissions arrive at the receiver within a backoff slot duration \cite{wanet.roy_etal09ToN-PCS}. Simultaneous transmissions by nodes that cannot sense each other's transmissions lead to \textit{asynchronous} (or hidden node) collisions at a receiver. In our network model, there cannot be packet losses due to asynchronous hidden node collisions (A2). However, there can be packet losses due to synchronous collisions, e.g., A0 and A3 imply that simultaneous transmissions in the same cell are always lost due to (synchronous) collisions.

\section{An Analytical Model for Multi-Cell WLANs with Arbitrary Cell Topology}
\label{sec:analysis-multicell-arbitrary-cell-topology}

%In this section, we develop a cell-level model for WLANs that satisfy A0-A5. Two important cases in which the PBD condition (A1) does not strictly hold (but our analytical model provides good approximations) will be briefly considered in Section \ref{subsec:when-PBD-does-not-hold}.

In this section, we develop a cell-level model for WLANs that satisfy A0-A5. We index the cells by positive integers $1, 2, \ldots, N$, where $N$ denotes the number of cells. Let $\mathcal{N} = \{1, 2, \ldots, N\}$ denote the set of cells. Let $M$ denote the number of non-overlapping channels, and let $\mathcal{C} = \{1, 2, \ldots, M\}$ denote the set of channels. Let $\bmath{c} = (c_1, c_2, \ldots, c_N)$ denote the \textit{channel assignment} where, $\forall i \in \mathcal{N}$, $c_i \in \mathcal{C}$ denotes the channel assigned to Cell-$i$. 

We construct a \textit{contention graph} $\mathcal{G}(\bmath{c})$ as follows. Every cell in the network is represented by a vertex in $\mathcal{G}(\bmath{c})$. There exists an edge between two vertices in $\mathcal{G}(\bmath{c})$ if and only if the corresponding cells are completely dependent. Owing to the PBD condition, any two cells are either independent or completely dependent. Moreover, two cells are completely dependent if the distance between the respective APs is less than $R_{cs}$ and the two cells operate on the same channel. Thus, given the locations of the APs and a channel assignment $\bmath{c}$, $\mathcal{G}(\bmath{c})$ can be constructed, and our model applies to any $\mathcal{G}(\bmath{c})$. To simplify notations, we assume the channel assignment $\bmath{c}$ to be given and fixed, and henceforth, we denote the contention graph $\mathcal{G}(\bmath{c})$ (for the given channel assignment $\bmath{c}$) by $\mathcal{G}$. Two cells whose corresponding vertices in the contention graph $\mathcal{G}$ are joined by an edge will be called \textit{neighbors}. Let $\mathcal{N}_i \; (\subset \mathcal{N})$ denote the set of neighboring cells of Cell-$i$ $(i \in \mathcal{N})$. By definition, $i \notin \mathcal{N}_i$.

\begin{figure}[tb]
  \centering \
  \begin{minipage}{1.5cm}
    \begin{center}
      \includegraphics[scale=0.2]{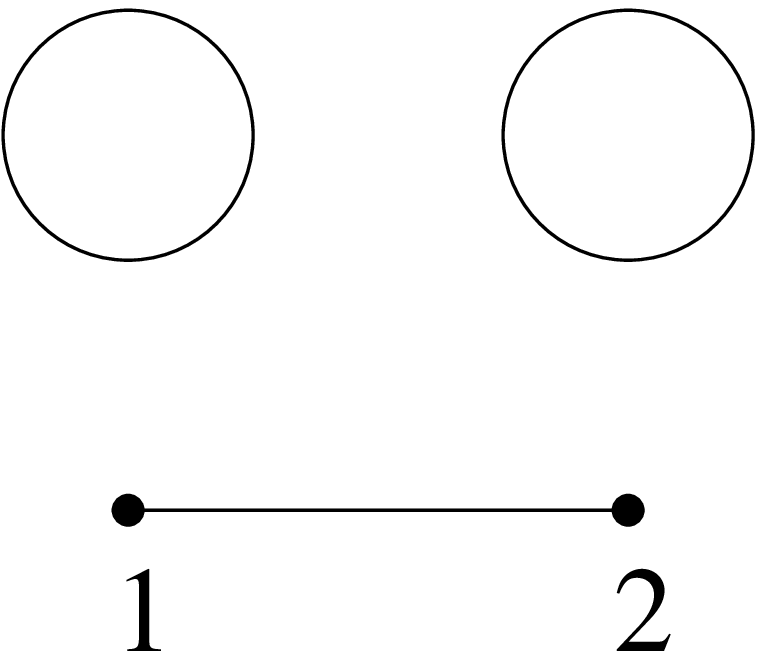}
      \subfigure[]{\label{fig:twoCell}}
    \end{center}
  \end{minipage}
%  \hfill
  \hspace{1cm}
  \begin{minipage}{2.5cm}
    \begin{center}
      \includegraphics[scale=0.2]{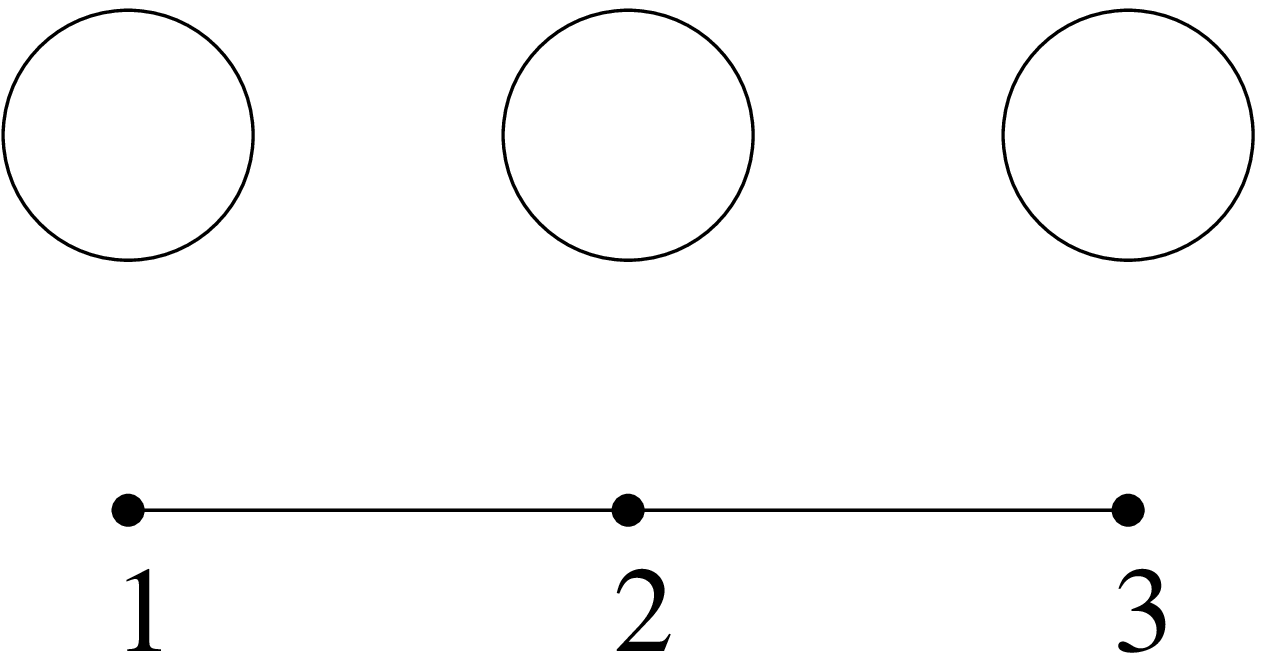}
      \subfigure[]{\label{fig:threeCellLinear}}
    \end{center}
  \end{minipage}
%  \hfill
  \hspace{2cm}
  \begin{minipage}{2.75cm}
    \begin{center}
      \includegraphics[scale=0.2]{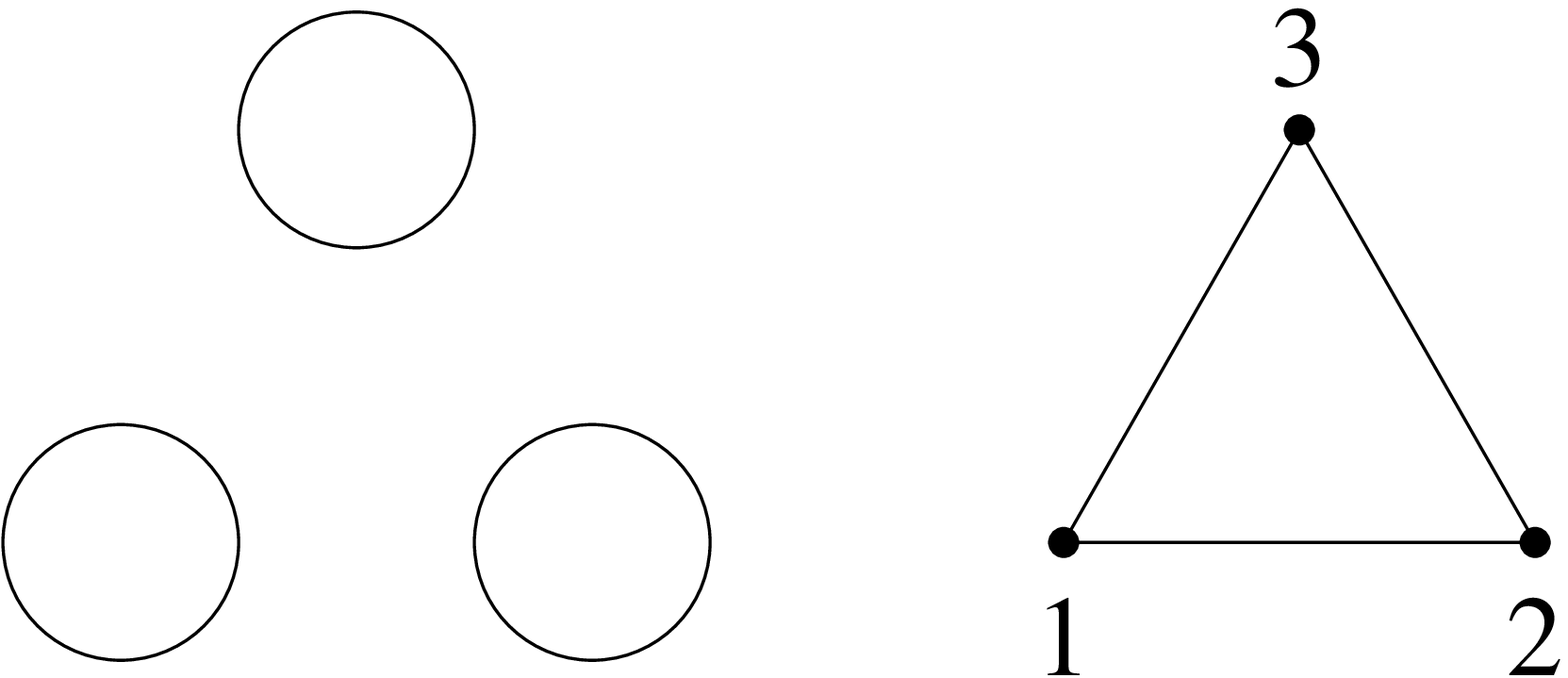}
      \subfigure[]{\label{fig:threeCellSingleCD}}
    \end{center}
  \end{minipage}
%  \hfill
  \hspace{1.5cm}
  \begin{minipage}{2.75cm}
    \begin{center}
      \includegraphics[scale=0.125]{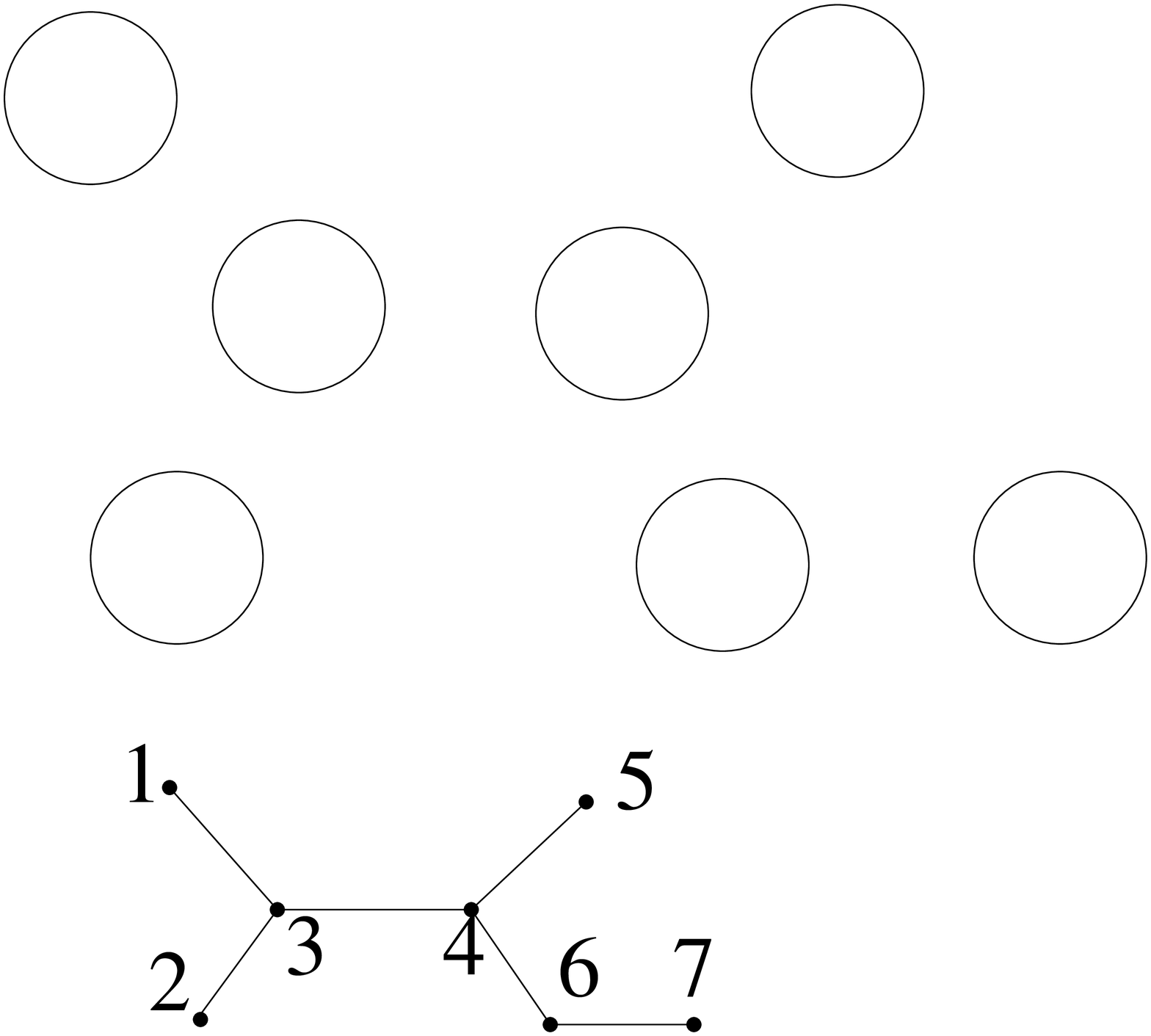}
      \subfigure[]{\label{fig:sevenCellArbitrary}}
    \end{center}
  \end{minipage}
  \hfill
  \caption{Examples of multi-cell systems: (a) two co-channel cells
    with a single contention domain, (b) three linearly placed
    co-channel cells with two contention domains, (c) three co-channel
    cells with a single contention domain, and (d) seven co-channel
    cells with an arbitrary cell topology and seven contention
    domains. The contention graphs have also been shown.}
%\vspace{-2mm}
\end{figure}

We first develop our model for saturated nodes (Section \ref{subsec:saturated-case}), and then extend to TCP-controlled long-lived downloads (Section \ref{subsec:TCP-traffic}). The case of TCP-controlled short-lived downloads is covered in Section \ref{sec:short-lived}. We develop a generic model and demonstrate the accuracy of our model by comparing with NS-2 simulations of a few example networks given in Figure \ref{fig:twoCell}-\ref{fig:sevenCellArbitrary}. Note that in Figure \ref{fig:twoCell}-\ref{fig:sevenCellArbitrary}, by a \textit{``contention domain''} we mean a \textit{maximal clique} of $\mathcal{G}$.

%\vspace{-2mm}

\subsection{Modeling with Saturated Nodes and UDP Traffic}
\label{subsec:saturated-case}

Consider the scenario where the nodes (i.e., the APs and the STAs) are \textit{saturated} or infinitely backlogged, and are transferring packets to one or more nodes in the same cell over UDP. Let $n_i$, $n_i \geq 2$, denote the number of nodes in Cell-$i$. Thus, Cell-$i$ consists of a saturated AP, AP-$i$, and $n_i - 1$ saturated STAs. The APs and their associated STAs exchange fixed size packets over UDP.

In general, nodes belonging to different cells can have different views of the network activity. Due to the PBD condition, however, nodes in the same cell have an identical view of the rest of the network. When one node senses the medium idle, so do the other nodes in the same cell, and we say that \textit{the cell is sensing the medium idle}. Since the nodes are saturated, whenever a cell senses the medium idle, all the nodes in the cell decrement their backoff counters per idle backoff slot that elapses in the \textit{local} medium of the cell, and we say that \textit{the cell is in backoff}. If the nodes were not saturated, a node with an empty transmission queue would not have been counting down during the ``medium idle'' periods and the number of \textit{contending} nodes would have been time-varying. With saturated nodes, however, nodes contend whenever they sense the medium idle, and the number of contending nodes in each cell remains constant. 

We say that \textit{a cell is transmitting when one or more nodes in the cell are transmitting}. When a cell starts transmitting, its neighbors (i.e., neighboring cells) sense the transmission within one backoff slot and they defer medium access. We then say that the neighbors are \textit{blocked} due to carrier sensing. When a cell is blocked, the backoff counters of all the nodes in the cell are \textit{frozen}. Thus, a cell can be in one of three states: (a) \textbf{Transmitting} (when at least one node in the cell is transmitting), (b) \textbf{Backoff} (when every node in the cell is decrementing its backoff counter per idle backoff slot), or (c) \textbf{Blocked} (when the backoff counters of all the nodes in the cell are frozen due to transmissions in some neighboring cell(s)).

%\subsubsection{A Two-stage Approach}
%\label{subsubsec:two-stage-approach}

We emphasize that \textit{the evolution of the network is partly synchronous and partly asynchronous}. Nodes in the same cell have an identical view of the rest of the network and they are synchronized. When two or more nodes in the same cell start transmitting together, a synchronous \textit{intra-cell} collision occurs. Two neighboring cells can start transmitting together (i.e., within a backoff slot) before they could sense each other's transmissions, resulting in synchronous \textit{inter-cell} collisions. For example, consider Figure \ref{fig:threeCellLinear} and suppose that Cell-1 and Cell-2 are counting down together. With positive probability, there would be simultaneous transmissions in both Cell-1 and Cell-2 leading to synchronous inter-cell collisions. Recall that, due to Assumption (A2), there cannot be asynchronous hidden node collisions in the network. However, blocking due to carrier sensing is, in general, asynchronous in nature. For example, consider Figure \ref{fig:threeCellLinear} and suppose that Cell-1 starts transmitting first and blocks Cell-2 (no later than a backoff slot). Since Cell-3 is independent of Cell-1, Cell-3 can start transmitting at any instant during the transmission in Cell-1. Thus, transmissions in Cell-1 and Cell-3 would overlap at the nodes in Cell-2 in an asynchronous manner keeping Cell-2 blocked. 

%In fact, the three-cell network in Figure \ref{fig:threeCellLinear} is the cell-level analogue of a three-link scenario known as the \textit{Flow In the Middle} (FIM) topology \cite{wanet.wang-kar05multihop,wanet.garetto_etal08starvation}.

%Modeling synchronous inter-cell collisions requires a discrete time slotted model. However, this would require a large state space since the cells change their states in an asynchronous manner due to asynchronous inter-cell blocking. 

To capture both the synchrony and the asynchrony, we extend the \textit{two-stage} approach of \cite{wanet.garetto_etal08starvation} from node-level to cell-level. In the first stage, we ignore inter-cell collisions and assume that inter-cell blocking due to carrier sensing is \textit{immediate}. We develop a continuous time model by extending the independent-set approach of \cite{wanet.boorstyn87multihop} from node-level to cell-level, and obtain the fraction of \textit{time} for which each cell is transmitting/blocked/in backoff. This requires careful choice of cell-level transition rates, which we obtain in \eqref{eqn:lambda_i-multicell-CTMC} and \eqref{eqn:mu_i-multicell-CTMC}. In the second stage, we obtain the fraction of \textit{backoff slots} in which various subsets of neighboring cells can start transmitting together, and compute the collision probabilities in \eqref{eqn:gamma_i-multicell} by accounting for synchronous inter-cell collisions. We combine the above by a fixed-point equation and compute the throughputs using the solution of the fixed-point equation. We define the following notation: 

%(adopted from \cite{wanet.kumar_etal07new_insights}, which provides a generalization of the model by \cite{wanet.bianchi00performance}):

\vspace{1mm}

\noindent $\beta_i :=$ \parbox[t] {7.5cm} {(transmission) attempt probability (over the backoff slots) of the nodes in Cell-$i$}

\vspace{1mm}

\noindent $\gamma_i :=$ \parbox[t] {7.6cm} {conditional collision probability of the nodes in Cell-$i$ (conditioned on an attempt being made)}

\vspace{1mm}

Thus, at every backoff slot boundary, every node in Cell-$i$ starts transmitting with probability $\beta_i$, i.e., the attempt process of each node is a Bernoulli process over the backoff slots. As in \cite{wanet.kumar_etal07new_insights}, $\beta_i$ is related to $\gamma_i$ by \begin{equation} 
\label{eqn:G_gamma_i}
\beta_i = G(\gamma_i) := \frac{1 + \gamma_i \dots + \gamma_i^K}{b_0 + \gamma_i b_1 \dots + \gamma_i^k b_k + \dots + \gamma_i^K b_K}, 
\end{equation} 
where $K$ denotes the \textit{retransmit limit} and $b_k$, $0 \leq k \leq K$, denotes the mean backoff sampled after the $k^{th}$ collision (for the same packet). The \textit{backoff parameters} $K$ and $b_k$, $0 \leq k \leq K$, are fixed in the DCF and their values depend on the PHY layer being used.

%\begin{remarks}
%\label{rmk:eqn:G_gamma}
%Equation \eqref{eqn:G_gamma_i} models the adaptation of the contention
%window of a node in response to failed transmissions. The algorithm by
%which a node adapts its contention window in response to failed
%transmissions is internal to the node, and does not depend on whether
%the network is a single cell or a multi-cell infrastructure
%WLAN. Thus, even though Equation \eqref{eqn:G_gamma_i} has been obtained
%in \cite{wanet.kumar_etal07new_insights} in the context of single
%cells, it can be applied in any network setting. \hfill \qed
%\end{remarks}

\subsubsection{The First Stage}
\label{subsubsec:first-stage}

When Cell-$i$ and some (or all) of its neighboring cells are in backoff, they contend until one of the cells, say, Cell-$j$, ${j \in \mathcal{N}_i \cup \{i\}}$, starts transmitting. Since we ignore inter-cell collisions in the first stage, the possibility of two or more neighboring cells starting transmission together is ruled out. When Cell-$i$ starts transmitting, it gains control over its local medium by \textit{immediately} blocking its neighbors that are not yet blocked. We assume that the time until Cell-$i$ goes from the backoff state to the transmitting state is exponentially distributed with mean $1/\lambda_i$.  

The \textit{``activation rate''} $\lambda_i$ of Cell-$i$ is given by 
\begin{equation} 
\label{eqn:lambda_i-multicell-CTMC}
\quad \quad \quad \quad \quad \quad \lambda_i = \frac{1 - (1 - \beta_i)^{n_i}}{\sigma}, 
\end{equation} 
where we recall that $n_i$ denotes the number of nodes in Cell-$i$, $\sigma$ denotes the duration of a backoff slot (in seconds), and $1 - (1 - \beta_i)^{n_i}$ is the probability that there is an attempt in Cell-$i$ per backoff slot.

\begin{remark}
\label{rmk:exponential-activation}
In \eqref{eqn:lambda_i-multicell-CTMC} we have converted the aggregate transmission attempt probability in a cell per (discrete) backoff \textit{slot} to an attempt rate over (continuous) backoff \textit{time}.  Our assumption of ``exponential time until transition from the backoff state to the transmitting state'' is the continuous time analogue of the assumption of ``geometric number of backoff slots until transmission attempt'' in the discrete time models of \cite{wanet.bianchi00performance} and \cite{wanet.kumar_etal07new_insights}. \hfill \qed
\end{remark}

\begin{remark}
\label{rmk:compare-garetto}
\cite{wanet.garetto_etal08starvation} use an unconditional activation rate $\lambda$ over all times as well as a conditional activation rate $g$ over the backoff times. We use a single activation rate $\lambda$ which is conditional on being in the backoff state. Our modified approach with a single conditional activation rate can also be applied to simplify the node- and link-level model of \cite{wanet.garetto_etal08starvation}. \hfill \qed
\end{remark}

%Since our model could capture the performance well, it is clear that the unconditional activation rate is not necessary. 

When Cell-$i$ transmits, its neighbor cells remain blocked (due to Cell-$i$) until Cell-$i$'s transmission finishes and an idle DIFS period elapses. If the transmission is a success (resp.~an intra-cell collision) in Cell-$i$, then Cell-$i$'s neighbors remain blocked for a \textit{success time} $T_s$ (resp.~a \textit{collision time} $T_c$). In the Basic Access (resp.~RTS/CTS) mechanism, $T_s$ corresponds to the time DATA-SIFS-ACK-DIFS (resp.~RTS-SIFS-CTS-SIFS-DATA-SIFS-ACK-DIFS) and $T_c$ corresponds to the time DATA-DIFS (resp.~RTS-DIFS). $T_s$ and $T_c$ can be computed using the protocol parameters and mean DATA payload size $L$. We denote the mean duration for which Cell-$i$'s neighbors remain blocked due to a transmission in Cell-$i$ by $1/\mu_i$, where \begin{equation}
\label{eqn:mu_i-multicell-CTMC}
\quad \quad \quad \quad \frac{1}{\mu_i} = p_{succ,i} \cdot T_s + (1-p_{succ,i}) \cdot T_c,
%\frac{1}{\mu_i} = \left( \frac{n_i \beta_i (1 - \beta_i)^{n_i - 1}}{1
%  - (1 - \beta_i)^{n_i}} \right) \cdot T_s + \left(1 - \frac{n_i
%  \beta_i (1 - \beta_i)^{n_i - 1}}{1 - (1 - \beta_i)^{n_i}} \right)
%\cdot T_c.
\end{equation} 
where $p_{succ,i} = \frac{n_i \beta_i (1 -\beta_i)^{n_i - 1}}{1 - (1 - \beta_i)^{n_i}}$ is the probability that the transmission in Cell-$i$ is successful given that there is a transmission in Cell-$i$. We take the activity periods of Cell-$i$ to be i.i.d. exponential random variables with mean $1/\mu_i$. 

Our approximation of exponential activity periods is justified by a well-known insensitivity result (see \cite[Theorem 1]{queueing.whittle85partial-balance-insensitivity} and \cite{wanet.boorstyn87multihop}) and the detailed balance equations satisfied by our model (see Equation \eqref{eqn:detailed-balance-multicell}). 

Due to carrier sensing, at any point of time, only a set ${\mathcal{A} \; (\subset \mathcal{N})}$ of mutually independent \textit{cells} can be transmitting together, i.e., $\mathcal{A}$ must be an \textit{independent set} (of vertices) of $\mathcal{G}$. From $\mathcal{G}$, we can determine the set of cells $\mathcal{B}_{\mathcal{A}}$ that get blocked due to $\mathcal{A}$, and the set of cells $\mathcal{U}_{\mathcal{A}}$ that remain in backoff, i.e., the set of cells in which nodes can continue to decrement their backoff counters. Note that $\mathcal{A}$, $\mathcal{B}_{\mathcal{A}}$ and $\mathcal{U}_{\mathcal{A}}$ form a partition of $\mathcal{N}$. 

\textit{We take $\mathcal{A}(t)$, i.e., the set $\mathcal{A}$ of cells that are transmitting at time $t$, as the state of the multi-cell system at time $t$}. Due to the exponential approximations of the distributions of the time to activation and the activity periods, at any time $t$, the rate of transition to the next state is completely determined by the current state $\mathcal{A}(t)$. For example, Cell-$j$, $j \in \mathcal{U}_{\mathcal{A}}$, joins the set $\mathcal{A}$ (and its neighboring cells that are also in $\mathcal{U}_{\mathcal{A}}$ join the set $\mathcal{B}_{\mathcal{A}}$) at a rate $\lambda_j$. Similarly, Cell-$i$, $i \in \mathcal{A}$, leaves the set $\mathcal{A}$ (and its neighboring cells that are blocked only due to Cell-$i$ leave the set $\mathcal{B}_{\mathcal{A}}$) to join the set $\mathcal{U}_{\mathcal{A}}$ at a rate $\mu_i$. We conclude that, \textit{the process $\{\mathcal{A}(t), t \geq 0\}$ has the structure of a Continuous Time Markov Chain} (CTMC). 

\begin{figure}[tb]
  \centering \
  \begin{minipage}{3.5cm}
    \begin{center}
      \includegraphics[scale=0.225]{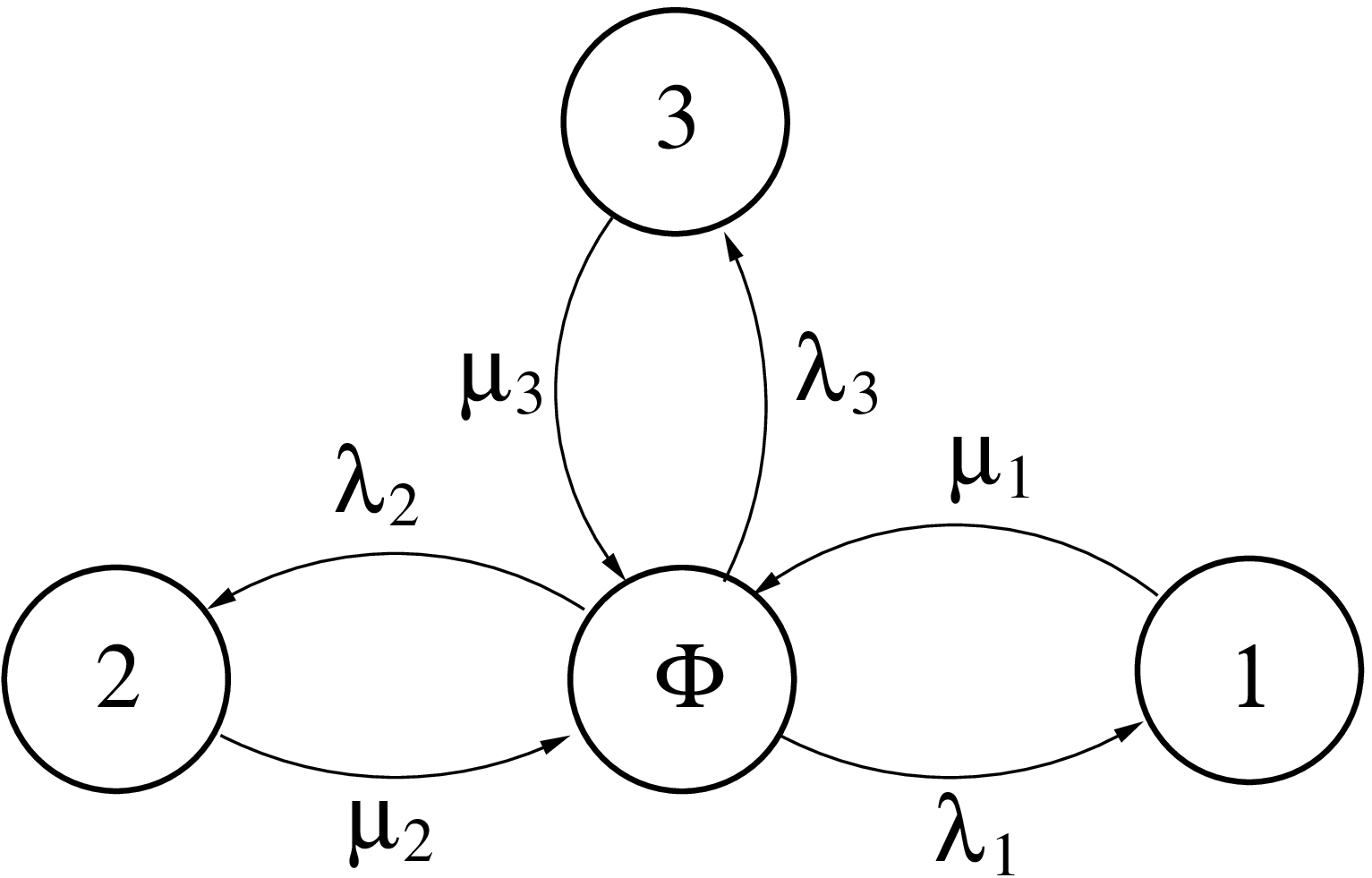}
      \subfigure[]{\label{fig:CTMCthreeCellSingleCD}}
    \end{center}
  \end{minipage}
  \hspace{0.5cm}
  \begin{minipage}{3.5cm}
    \begin{center}      \includegraphics[scale=0.225]{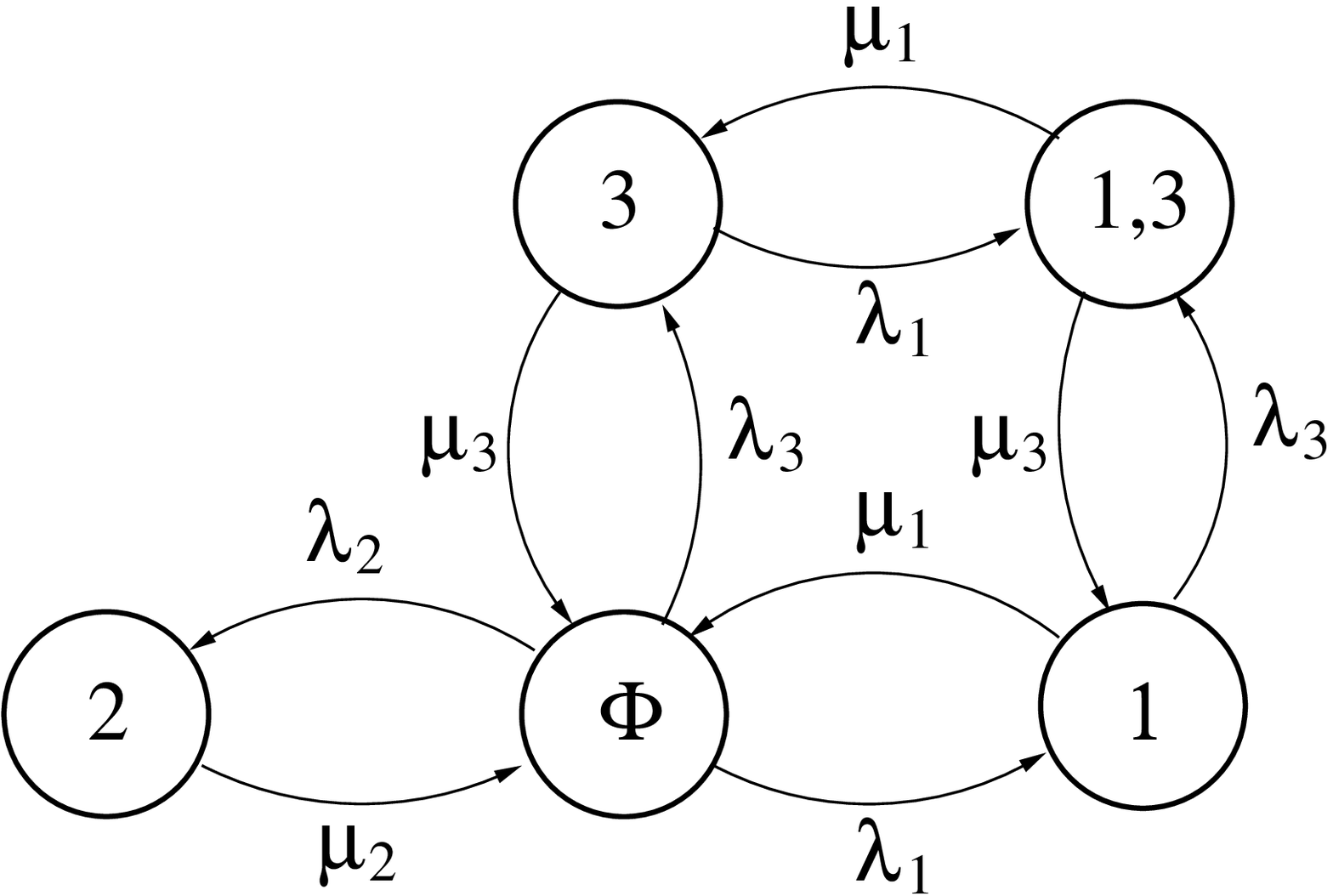}
      \subfigure[]{\label{fig:CTMCthreeCellLinear}}
    \end{center}
  \end{minipage}
  \caption{The CTMCs describing the cell-level contention for example
    networks in (a) Figure \ref{fig:threeCellSingleCD}, and (b) Figure
    \ref{fig:threeCellLinear}.}
%\vspace{-1mm}
\end{figure}

The CTMC $\{\mathcal{A}(t), t \geq 0\}$ has a finite number of states, and it is irreducible as long as the $\lambda_i$'s and the $\mu_i$'s are non-zero, which is the case when the contention windows and the packet payloads are finite. Then, the CTMC $\{\mathcal{A}(t), t \geq 0\}$ is \textit{stationary} and \textit{ergodic}. The set of all possible independent sets of $\mathcal{G}$ which constitutes the state space of the CTMC $\{\mathcal{A}(t), t \geq 0\}$ is denoted by $\bmath{\mathcal{A}}$. The state when none of the cells is transmitting will be denoted by $\Phi$, where $\Phi$ denotes the \textit{empty set}. For a given contention graph $\mathcal{G}$, $\bmath{\mathcal{A}}$ can be determined. For the topology given in Figure \ref{fig:threeCellSingleCD} we have $\bmath{\mathcal{A}} = \{\Phi,\{1\},\{2\},\{3\}\}$, and for the topology given in Figure \ref{fig:threeCellLinear}, we have $\bmath{\mathcal{A}} = \{\Phi,\{1\},\{2\},\{3\},\{1,3\}\}$. The corresponding CTMCs are given in Figures \ref{fig:CTMCthreeCellSingleCD} and \ref{fig:CTMCthreeCellLinear}.

It can be checked that the transition structure of the CTMC $\{\mathcal{A}(t), t \geq 0\}$ satisfies the ``Kolmogorov Criterion'' for reversibility \cite{theory.kelly79reversibility}. Hence, the stationary probability distribution $\pi(\mathcal{A}), \mathcal{A} \in \bmath{\mathcal{A}}$, satisfies the \textit{detailed balance equations}, $\forall i \in \mathcal{U}_{\mathcal{A}}$, 
\begin{equation}
\label{eqn:detailed-balance-multicell}
\quad \quad \quad \quad \quad \quad \quad \pi(\mathcal{A})\lambda_i = \pi(\mathcal{A} \cup \{i\})\mu_i.
\end{equation}

Define the \textit{``access intensity''} $\rho_i$ of Cell-$i$ by 
\begin{equation}
\label{eqn:varrho-definition}
\quad \quad \quad \quad \quad \quad \quad \quad \quad \quad \rho_i := \frac{\lambda_i}{\mu_i}.
\end{equation} 
Due to reversibility, the stationary distribution of $\{\mathcal{A}(t), t \geq 0\}$ has the following product-form, $\forall \mathcal{A} \in \bmath{\mathcal{A}}$, 
\begin{equation}
\label{eqn:stationary-probabilities}
\quad \quad \quad \quad \quad \quad \quad \pi(\mathcal{A}) = \left( \prod_{i \in \mathcal{A}}\rho_i \right) \pi(\Phi), 
\end{equation} 
where $\pi(\Phi)$ denotes the stationary probability that none of the cells is transmitting. $\pi(\Phi)$ can be obtained by applying the normalization equation 
\begin{equation}
\label{eqn:normalization-equation}
\quad \quad \quad \quad \quad \quad \quad \quad \quad \sum_{\mathcal{A} \in \bmath{\mathcal{A}}} \pi(\mathcal{A}) = 1.
\end{equation}

Combining \eqref{eqn:stationary-probabilities} and \eqref{eqn:normalization-equation}, we obtain, $\forall \mathcal{A} \in \bmath{\mathcal{A}}$, the stationary probability $\pi(\mathcal{A})$ that the particular set $\mathcal{A}$ of cells is transmitting by 
\begin{equation}
\label{eqn:stationary-probabilities-complete}
\quad \quad \quad \quad \quad \quad \pi(\mathcal{A}) = \frac{\left( \prod_{i \in \mathcal{A}}\rho_i
  \right)}{\sum_{\mathcal{A} \in \bmath{\mathcal{A}}} \left( \prod_{j
    \in \mathcal{A}}\rho_j \right)}, 
\end{equation} 
where we recall that a product $\prod$ over an empty index set is taken to be equal to 1.

\subsubsection{The Second Stage}
\label{subsubsec:second-stage}

In the first stage analysis in Section \ref{subsubsec:first-stage}, we had ignored inter-cell collisions. We now compute the collision probabilities $\gamma_i$'s accounting for inter-cell collisions.  Note that $\gamma_i$ is conditional on an attempt being made by a node in Cell-$i$. Hence, to compute $\gamma_i$, we focus only on those states in which Cell-$i$ can attempt. Clearly, Cell-$i$ can attempt in State-$\mathcal{A}$ iff it is in back-off in State-$\mathcal{A}$, i.e., iff $i \in \mathcal{U}_{\mathcal{A}}$. In all such states a node in Cell-$i$ can incur intra-cell collisions due the other nodes in Cell-$i$. Furthermore, some (or all) of Cell-$i$'s neighbors might also be in back-off in State-$\mathcal{A}$.  If a neighboring cell, say, Cell-$j$, $j \in \mathcal{N}_i$, is also in back-off in State-$\mathcal{A}$, i.e., if $j \in \mathcal{U}_{\mathcal{A}}$, then a node in Cell-$i$ can incur inter-cell collisions due to the nodes in Cell-$j$.  The collision probability $\gamma_i$ is then given by, $\forall i \in \mathcal{N}$, 
\begin{equation}
\label{eqn:gamma_i-multicell}
\quad \quad \quad \quad \quad \gamma_i = \frac{ \sum_{\mathcal{A} \in \bmath{\mathcal{A}} \; : \; i
    \in \mathcal{U}_{\mathcal{A}}} \pi(\mathcal{A})
  \gamma_i(\mathcal{A})}{ \sum_{\mathcal{A} \in \bmath{\mathcal{A}} \;
    : \; i \in \mathcal{U}_{\mathcal{A}}} \pi(\mathcal{A})},
\end{equation} 
where 
\[\quad \quad \gamma_i(\mathcal{A}) = 1 - (1-\beta_i)^{n_i-1} \prod_{j \in \mathcal{N}_i \; : \; j \in \mathcal{U}_{\mathcal{A}}} (1-\beta_j)^{n_j}\] 
represents the (conditional) collision probability of the nodes in Cell-$i$ in State-$\mathcal{A}$. A derivation of \eqref{eqn:gamma_i-multicell} can be found in the Appendix at the end of the paper. We emphasize that \eqref{eqn:gamma_i-multicell} is different from the equation for synchronous collisions, Equation (10), of \cite{wanet.garetto_etal08starvation}.

%\cite{thesisPhdManoj2010}. 

%A formal derivation
%of Equation \eqref{eqn:gamma_i-multicell} can be found in Appendix A of
%\cite{wanet.manoj-anurag10multicellTechReportWithShortTCP}.

\subsubsection{Fixed Point Formulation}
\label{subsubsec:fixed-point-formulation}

The $\gamma_i$'s can be written as functions of the $\beta_i$'s by applying \eqref{eqn:lambda_i-multicell-CTMC}-\eqref{eqn:gamma_i-multicell}. Let, $\forall i \in \mathcal{N}$, 
\[\quad \quad \quad \quad \quad \quad \quad \gamma_i = \Gamma_i(\beta_1, \beta_2, \ldots, \beta_N)\] 
denote this functional dependence. Also, $\forall i \in \mathcal{N}$, \eqref{eqn:G_gamma_i} provides the functional dependence of $\beta_i$ on $\gamma_i$ as $\beta_i = G(\gamma_i)$. Using the notation $\bmath{\beta} = (\beta_1, \beta_2, \ldots, \beta_N)$, we obtain an $N$-dimensional fixed point equation in terms of $\bmath{\beta}$ given by 
\begin{equation}
\label{eqn:N-dim-multi-cell-fixed-point}
\quad \quad \quad \bmath{\beta} = (G(\Gamma_1(\bmath{\beta})), G(\Gamma_2(\bmath{\beta})), \ldots, G(\Gamma_N(\bmath{\beta}))),
\end{equation} 
where we recall that $N$ is the total number of cells. This $N$-dimensional fixed point equation can be numerically solved to obtain the attempt probabilities $\beta_i$'s. Once the $\beta_i$'s are known, the stationary state probabilities $\pi(\mathcal{A})$'s and the collision probabilities $\gamma_i$'s can also be obtained by applying \eqref{eqn:lambda_i-multicell-CTMC}-\eqref{eqn:gamma_i-multicell}. In all of the cases that we have considered, the fixed point iterations were observed to converge to the same solutions irrespective of starting points.

\subsubsection{Computation of Saturation Throughputs}
\label{subsubsec:computing-cell-throughputs}

Let $x_i(\mathcal{G})$ denote the fraction of time for which Cell-$i$ is \textit{not} blocked in a multi-cell network with contention graph $\mathcal{G}$. Thus, $x_i(\mathcal{G})$ is the fraction of time when Cell-$i$ is either transmitting ($i \in \mathcal{A}$) or in backoff ($i \in \mathcal{U}_{\mathcal{A}}$), and hence, $\forall i \in \mathcal{N}$, 
\begin{equation}
\label{eqn:fraction-of-channel-time}
\quad \quad \quad \quad \quad x_i(\mathcal{G}) = \sum_{\mathcal{A} \in \bmath{\mathcal{A}}\; : \; i \in \mathcal{A} \cup \mathcal{U}_{\mathcal{A}}} \pi(\mathcal{A}), 
\end{equation}
where we recall that $\bmath{\mathcal{A}}$ depends on $\mathcal{G}$. 

Let $\Theta_i^{sat}(\mathcal{G})$ denote the aggregate throughput in packets/sec of (the $n_i$ saturated nodes in) Cell-$i$ in the multi-cell network with contention graph $\mathcal{G}$. Let $\Theta_{n,singlecell}^{sat}$ denote the aggregate throughput in packets/sec of a single \textit{isolated} cell containing $n$ saturated nodes. Note that $\Theta_{n,singlecell}^{sat}$ can be computed as in \cite{wanet.bianchi00performance} or \cite{wanet.kumar_etal07new_insights}. We approximate $\Theta_i^{sat}(\mathcal{G})$ by 
\begin{equation}
\label{eqn:Theta-multicell}
\quad \quad \quad \quad \quad \Theta_{i}^{sat}(\mathcal{G}) \approx x_i(\mathcal{G}) \cdot \Theta_{n_i,singlecell}^{sat}
\end{equation} 
and $\Theta_i^{sat}(\mathcal{G})$ divided by $n_i$ gives the per node throughput $\theta_i^{sat}(\mathcal{G})$ in Cell-$i$, i.e., $\theta_i^{sat}(\mathcal{G}) = \Theta_i^{sat}(\mathcal{G})/n_i$.

%(packets/sec).

If Cell-$i$ is indeed an isolated cell, then we have $x_i(\mathcal{G}) = 1$ and $\Theta_i^{sat}(\mathcal{G}) = \Theta_{n_i,singlecell}^{sat}$. However, in general, Cell-$i$ remains blocked (by other cells' activity) for a fraction of time $1-x_i(\mathcal{G})$, and hence, $x_i(\mathcal{G})$ appears as the reduction factor in \eqref{eqn:Theta-multicell}. The approximation in \eqref{eqn:Theta-multicell} is explained as follows. If we could ignore the time wasted in inter-cell collisions, then the times during which Cell-$i$ is not blocked would consist of only of the backoff slots and the activities of Cell-$i$ by itself. However, in general, a fraction of the medium time is wasted in inter-cell collisions. Thus, the single cell throughput $\Theta_{n_i,singlecell}^{sat}$ of Cell-$i$ is only an approximation of the aggregate throughput of Cell-$i$, over the times during which it is not blocked, and $\Theta_{n_i,singlecell}^{sat}$ multiplied by $x_i(\mathcal{G})$ provides an approximation of the aggregate throughput $\Theta_i^{sat}(\mathcal{G})$ of Cell-$i$ in the multi-cell network. In Section \ref{sec:results} we show that our analytical results (that are based on \eqref{eqn:Theta-multicell}) are quite accurate when compared with simulations.

%\subsubsection{Complexity of the Model}
%\label{subsubsec:model-complexity}

\begin{remark}[Computational Complexity]
\label{rmk:model-complexity}
In general, the complexity of obtaining the state space $\bmath{\mathcal{A}}$ by searching for all possible independent sets $\mathcal{A}$, grows exponentially with the number of vertices in the contention graph \cite{wanet.kershenbaum-etal87complex}. For realistic topologies, where connectivity in the contention graph is related to distances between the nodes in the physical network, efficient computation of $\bmath{\mathcal{A}}$ is possible up to several hundred vertices in the contention graph \cite{wanet.kershenbaum-etal87complex}. Thus, a cell-level model is extremely useful in analyzing large-scale WLANs with hundreds of cells, since, unlike a node-level model, each vertex in the contention graph now represents a cell. \hfill \qed

\end{remark}

\subsection{Extension to TCP-controlled Long-lived Downloads}
\label{subsec:TCP-traffic}

We now extend the analysis of Section \ref{subsec:saturated-case} to the case when STAs download long files via their respective APs from a \textit{local} server using TCP connections. Our extension to TCP-controlled long-lived flows is based on the analytical model for multi-cell WLANs with saturated nodes, developed in Section \ref{subsec:saturated-case}, and the \textit{equivalent saturated model} of \cite{wanet.bruno08TCPeqvSatModel} for TCP-WLAN interaction in a single cell. The model proposed by \cite{wanet.bruno08TCPeqvSatModel} has been shown to be quite accurate under the following assumptions:

\begin{itemize}

\item [Aa] The TCP source agents reside in a \textit{local server} which is connected with the AP by a relatively fast wireline LAN (which is our network setting) such that the AP in the WLAN is the bottleneck for every TCP connection.

\item [Ab] Every STA has a \textit{single} long-lived TCP connection.

\item [Ac] There are no packet losses due to buffer overflow.

\item [Ad] The TCP timeouts are set large enough to avoid timeout expiration due to Round Trip Time (RTT) fluctuations.

\item [Ae] The delayed ACK mechanism has been disabled.

\item [Af] The TCP connections have equal maximum windows.

\end{itemize}

%We keep the above assumptions in this section and also for the case of TCP-controlled short-lived flows in Section \ref{sec:short-lived}.

We assume Aa-Af in this section. Under Aa-Af, \cite{wanet.bruno08TCPeqvSatModel} propose to model a single cell having an AP and an \textit{arbitrary} number of STAs (with long-lived TCP connections) by an \textit{equivalent saturated network} which consists of a saturated AP and a \textit{single} saturated STA. ``This equivalent saturated model greatly simplifies the modeling problem since the TCP flow control mechanisms are now implicitly hidden and the aggregate throughput can be computed using the saturation analysis \cite{wanet.bruno08TCPeqvSatModel}.'' 

Using the equivalent saturated model of \cite{wanet.bruno08TCPeqvSatModel}, the analysis of Section \ref{subsec:saturated-case} can be directly applied to TCP-controlled long file downloads by taking $n_i = 2, \forall i \in \mathcal{N}$, i.e., \textit{by assuming two saturated nodes per cell regardless of how many STAs are actually present in the cells}. Of the two saturated nodes, one represents a saturated AP and the other represents an equivalent saturated STA.

\begin{remark}
\label{rmk:TCP-up-down-traffic}
In TCP-controlled downloads, APs send TCP DATA packets to their associated STAs and STAs send TCP ACK packets to their respective APs. Note that, both TCP DATA and TCP ACK packets are DATA units from the point of view of the MAC. Thus, to properly account for TCP DATA and TCP ACK transfers, we take the size of the MAC data units of one of the saturated nodes to be $L_{TCP-DATA}$ and that of the other saturated node to be $L_{TCP-ACK}$, where $L_{TCP-DATA}$ (resp.~$L_{TCP-ACK}$) denotes the size of a TCP DATA (resp.~TCP ACK) packet plus the IP header. \hfill \qed
\end{remark}

\subsubsection{AP Throughputs with Long-lived Flows}
\label{subsubsec:computing-cell-throughputs-TCP-Long}

The throughput of an AP, by definition, is equal to the aggregate throughput obtained by all the users/STAs that are served by the AP. Let $\theta_{singlecell}^{AP-TCP}$ denote the throughput in packets/sec of the AP in a single isolated cell under long-lived downloads. We obtain $\theta_{singlecell}^{AP-TCP}$ by applying the equivalent saturated model of \cite{wanet.bruno08TCPeqvSatModel}. Clearly, half the successful transmissions must belong to the AP (no-delayed-ACKs). Thus, we must have 
\[\quad \quad \quad \quad \quad \theta_{singlecell}^{AP-TCP} = \Theta_{n,singlecell}^{sat}/2,\] 
with $\Theta_{n,singlecell}^{sat}$ computed by taking $n=2$, and a payload size of $(L_{TCP-DATA}+L_{TCP-ACK})/2$. The throughput, $\theta_i^{AP-TCP}(\mathcal{G})$, of the AP in Cell-$i$ in a multi-cell network with contention graph $\mathcal{G}$, is given by 
\begin{equation}
\label{eqn:theta-i-AP-TCP}
\quad \quad \quad \quad \quad \theta_i^{AP-TCP}(\mathcal{G}) \approx x_i(\mathcal{G}) \cdot \theta_{singlecell}^{AP-TCP}, 
\end{equation}
where $x_i(\mathcal{G})$ is computed by the analysis of Section \ref{subsec:saturated-case}, taking $n_i = 2, \forall i \in \mathcal{N}$. 

%(see Remark \ref{rmk:TCP-up-down-traffic}).

%We take $n_i = 2, \forall i \in \mathcal{N}$, i.e., we assume that there are two saturated nodes per cell regardless of how many STAs are actually present in the cells. 

\subsection{Large $\rho$ Regime}
\label{subsec:large-rho-regime}

For fast computation of the cell throughputs, we extend (from node-level to cell-level) an approach originally proposed by \cite{wanet.wang-kar05multihop} and later adopted by \cite{wanet.durvy09selfOrganization} and \cite{wanet.liew_etal09ICCback-of-the-envelope}. As in \cite{wanet.wang-kar05multihop}, we take $\rho_i = \rho$, $\forall i \in \mathcal{N}$, and then let $\rho \longrightarrow \infty$. We call this \textit{the infinite $\rho$ approximation}, and justify its applicability to our WLAN setting later in this subsection (see Example \ref{ex:justifying-infinite-rho}). 

Let $\eta(\mathcal{G})$ denote the number of Maximum Independent Sets (MISs) of vertices of $\mathcal{G}$. Let $\eta_i(\mathcal{G})$ denote the number of MISs of vertices of $\mathcal{G}$ to which Cell-$i$ belongs. Let $\alpha(\mathcal{G})$ denote the cardinality of an MIS of vertices of $\mathcal{G}$. For any graph $G$, $\alpha(G)$ is called the \textit{independence number} of $G$. 

Applying the infinite $\rho$ approximation to \eqref{eqn:stationary-probabilities-complete}, we observe that, as $\rho \longrightarrow \infty$, we have 
\[\quad \quad \quad \pi(\mathcal{A}) \longrightarrow \left\{ \begin{array}{cl} \displaystyle \frac{1}{\eta(\mathcal{G})} & \mbox{if $\mathcal{A}$ is an MIS of $\mathcal{G}$,} \\ 0 & \mbox{otherwise}. \end{array} \right.\] 
Thus, as $\rho \longrightarrow \infty$, only an MIS of cells can be transmitting at any point of time. Note that when an MIS of cells is transmitting, all other cells must be blocked. This implies that, as $\rho \longrightarrow \infty$, every cell is either transmitting or is blocked at all times, i.e., the probability that a cell is in backoff goes to 0. This essentially provides a physical interpretation for infinite $\rho$ approximation, namely, \textit{``the medium time wasted in contention via backoffs is negligible as compared to the time spent in transmissions.''}

Applying the infinite $\rho$ approximation to \eqref{eqn:fraction-of-channel-time}, we observe that, as $\rho \longrightarrow \infty$, we have 
\begin{equation}
\label{eqn:xi-infty}
\quad \quad \quad \quad \quad \quad \quad x_i(\mathcal{G}) \longrightarrow \frac{\eta_i(\mathcal{G})}{\eta(\mathcal{G})}.
\end{equation} 
The quantity 
\[\quad \quad \quad \quad \quad \quad \quad x_i(\mathcal{G}) = \frac{\Theta_i}{\Theta_{singlecell}}\] 
can be interpreted as the throughput of Cell-$i$, normalized with respect to the single cell throughput (recall \eqref{eqn:Theta-multicell} and \eqref{eqn:theta-i-AP-TCP}). Defining the \textit{normalized network throughput} $\bar{\Theta}(\mathcal{G})$ by 
\begin{equation}
\label{eqn:normalized-network-throughput}
\quad \quad \quad \quad \quad \quad \quad \bar{\Theta}(\mathcal{G}) := \sum_{i=1}^N x_i(\mathcal{G}),
\end{equation} 
we prove the following theorem.

\begin{theorem}
\label{thm:maximum-throughput}

As $\rho_i = \rho \longrightarrow \infty$ for all $i \in \mathcal{N}$, we have 
\[\quad \quad \quad \quad \quad \quad \quad \bar{\Theta}(\mathcal{G}) \longrightarrow \alpha(\mathcal{G}).\]

\end{theorem}

\textit{Proof:} By definition, we have $\bar{\Theta}(\mathcal{G}) = \sum_{i=1}^N x_i(\mathcal{G})$. Also, as $\rho_i = \rho \longrightarrow \infty$ for all $i \in \mathcal{N}$, we have $x_i(\mathcal{G}) \longrightarrow \frac{\eta_i(\mathcal{G})}{\eta(\mathcal{G})}$. Summing over all $i \in \mathcal{N}$, as $\rho_i = \rho \longrightarrow \infty$ for all $i \in \mathcal{N}$, we have $\sum_{i=1}^N x_i(\mathcal{G}) \longrightarrow \sum_{i=1}^N \frac{\eta_i(\mathcal{G})}{\eta(\mathcal{G})}$. Next we prove that $\sum_{i=1}^N \frac{\eta_i(\mathcal{G})}{\eta(\mathcal{G})} = \alpha(\mathcal{G})$. Consider the equivalent equation 
\[\quad \quad \quad \quad \quad \quad \sum_{i=1}^N \eta_i(\mathcal{G}) = \alpha(\mathcal{G}) \; \eta(\mathcal{G}),\] 
which can be proved by a counting argument as follows. Imagine the $\eta(\mathcal{G})$ MISs are written as $\eta(\mathcal{G})$ rows with each row listing the cell indices in the corresponding MIS. Since each row contains $\alpha(\mathcal{G})$ elements and there are $\eta(\mathcal{G})$ rows, the total number of elements is equal to $\alpha(\mathcal{G}) \; \eta(\mathcal{G})$ which is given by the right hand side. The total number of elements can also be counted as follows. First count how many times Cell-$i$, $i = 1, 2, \ldots, N$, appears. Clearly, Cell-$i$ appears exactly $\eta_i(\mathcal{G})$ times. Now summing up the counts $\eta_i(\mathcal{G})$'s we get the total number of elements which is given by the left hand side. Hence, Theorem \ref{thm:maximum-throughput} is proved. \hfill \qed

\begin{remark}
\label{rmk:maximum-network-throughput}
Recall that, $\alpha(\mathcal{G})$ is the cardinality of an MIS of $\mathcal{G}$. Thus, the maximum possible value which the normalized network throughput $\bar{\Theta}(\mathcal{G})$ can take is $\alpha(\mathcal{G})$. Theorem \ref{thm:maximum-throughput} says that, as $\rho_i = \rho \longrightarrow \infty$ for all $i \in \mathcal{N}$, the normalized network throughput $\bar{\Theta}(\mathcal{G})$ is automatically maximized in a distributed manner without any additional control. This property of the CSMA/CA protocol has been recently established by \cite{wanet.durvy09selfOrganization} in the context of an infinite linear chain of saturated nodes. In Theorem \ref{thm:maximum-throughput} we prove it for multi-cell WLANs satisfying the PBD condition with arbitrary contention graph $\mathcal{G}$ and with saturated nodes as well as for TCP-controlled long-lived downloads. \hfill \qed
\end{remark}

%In Section \ref{sec:results} we provide simulation results in support of Theorem \ref{thm:maximum-throughput}. 

\begin{figure}[tb]
\centering
\begin{minipage}{8.25cm}
\begin{center}
\includegraphics[scale=0.85]{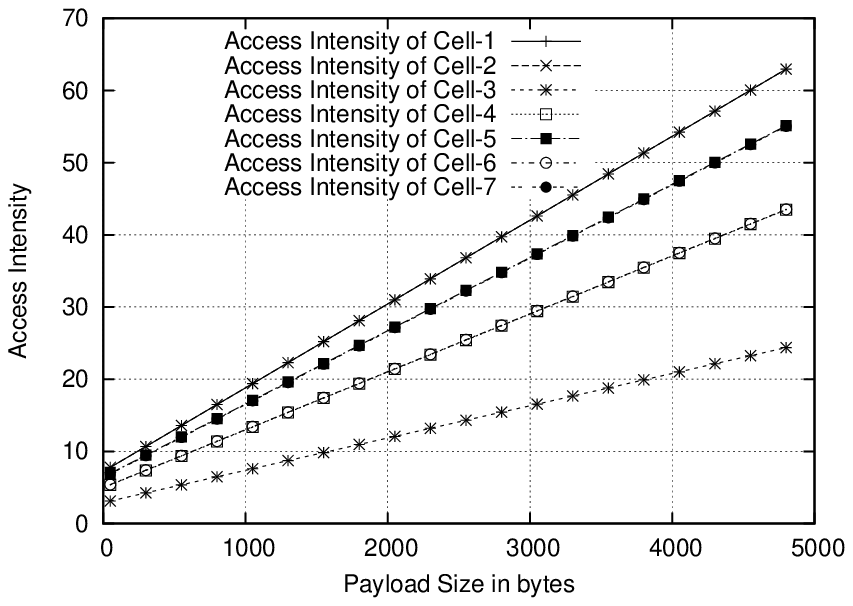}
\caption{Variation of access intensities with MAC payload size: for the scenario in Figure \ref{fig:sevenCellArbitrary} with 802.11b default parameter settings, 11 Mbps data and control rate, and 10 saturated nodes in each cell. \label{fig:rhoVsPayload}}
\end{center}
\end{minipage}
\hfill
\begin{minipage}{8.25cm}
\begin{center}
\includegraphics[scale=0.85]{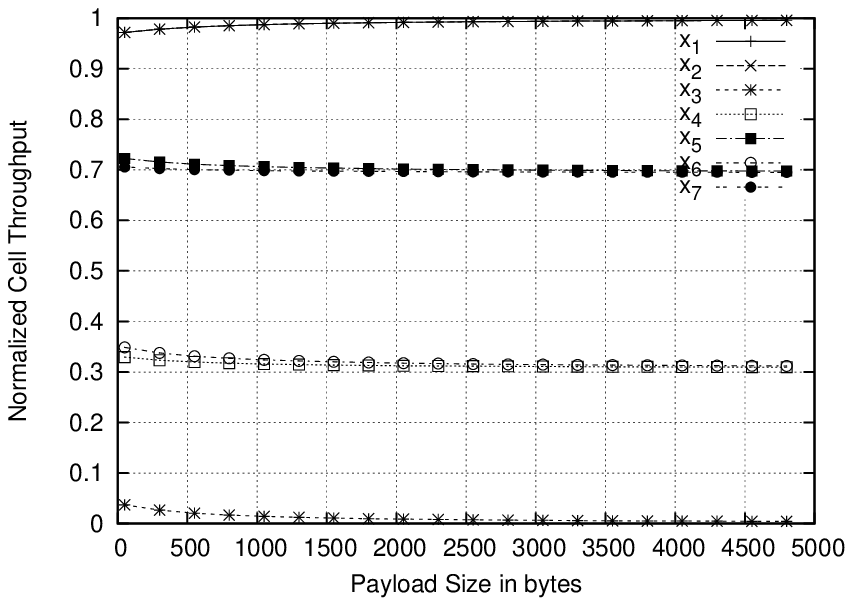}
\caption{Variation of normalized cell throughputs with MAC payload size: 
for the scenario in Figure \ref{fig:sevenCellArbitrary} with 802.11b default parameter settings, 11 Mbps rates, and 10 saturated nodes in each cell.\label{fig:xVsPayload}}
\end{center}
\end{minipage}
%\vspace{-2mm}
\end{figure}

We now justify the applicability of the infinite $\rho$ approximation. We provide an example to argue that \textit{realistic access intensities are indeed ``large'' in the sense that the values of the normalized cell throughputs with realistic access intensities are very close to their values with $\rho \longrightarrow \infty$}. 

\begin{example}
\label{ex:justifying-infinite-rho}
Consider the scenario in Figure \ref{fig:sevenCellArbitrary} with 802.11b default backoff parameter settings, 11 Mbps data and control rates, and 10 saturated nodes in each cell. Since, it is difficult to control $\rho$ directly, we examine the variation of the $x_i$'s with increase in $\rho$ in two steps. First, notice in Figure \ref{fig:rhoVsPayload} that the access intensities increase monotonically as the MAC payload size increases. Thus, it is reasonable to infer the variation of the $x_i$'s with $\rho$ by examining the variation of the $x_i$'s with the MAC payload size. Second, notice in Figure \ref{fig:xVsPayload} that, above a payload size, say, of 500 bytes, the $x_i$'s are very close to their values with $\rho \rightarrow \infty$, i.e., they are very close to the values $x_1 = x_2 = 1, x_3 = 0, x_4 = x_6 = 1/3, x_5 = x_7 = 2/3$, which are predicted by \eqref{eqn:xi-infty}. Figures \ref{fig:rhoVsPayload} and \ref{fig:xVsPayload} indicate that for access intensities as small as 5-15 (which correspond to the payload size of 500 bytes) the infinite $\rho$ approximation is already a very good approximation. \hfill \qed
\end{example}

\section{Short-lived Downloads}
\label{sec:short-lived}

In Section \ref{subsec:large-rho-regime} we showed that the infinite $\rho$ approximation can indeed be a very good approximation to predict cell throughputs at realistic access intensities. In this section, we apply the infinite $\rho$ approximation to develop a simple and accurate model under short-lived downloads, and to obtain predictions for mean flow transfer delays in each cell. We keep A0-A5 of Section \ref{sec:model-assumptions}, and also Aa-Af of Section \ref{subsec:TCP-traffic}, except Ab. As in \cite{wanet.bonald08multicellprocsharing}, applying the so-called \textit{time-scale separation} assumption, we model a multi-cell infrastructure WLAN under short-lived downloads as a network of state-dependent processor sharing queues. Our network model under A0-A5, Aa, and Ac-Af is a special case of the network model in \cite{wanet.bonald08multicellprocsharing} and our traffic model in this case is the same as that in \cite{wanet.bonald08multicellprocsharing}. The key difference lies in the service models.

%\vspace{-2mm}

\subsection{Traffic Model}
\label{subsubsec:traffic-model}

TCP-controlled short file downloads by users in Cell-$i$ and served by AP-$i$ are referred to as class-$i$ flows. Flows of class-$i$ arrive at AP-$i$ according to a Poisson process of rate $\nu_i$ (flows/sec). The flow arrival processes of different classes are independent. Flow sizes are i.i.d.~exponentially distributed random variables with mean $\EXP{V}$ (bits/flow).\footnote{It is possible to generalize to class-dependent flow size distributions, but we consider the same flow size distribution for all classes.} Flow sizes are independent of the arrival processes. Note that, flow sizes correspond to \textit{application-level} data.

\subsection{Network State}
\label{subsubsec:network-state}

Let $z_i$ denote the number of ongoing flows at AP-$i$ in Cell-$i$. We take 
\[\quad \quad \quad \quad \quad \quad \quad \bmath{z} = (z_1, z_2,\ldots, z_N)\] 
as the state of the network, where we recall that $N$ denotes the number of APs in the WLAN. Thus, the state space is $\mathbb{Z}_+^N$ where $\mathbb{Z}_+$ denotes the set of all non-negative integers.

%\vspace{-2mm}

\subsection{Service Model}
\label{subsubsec:our-service-model-short-flows}

We now develop a model for the service process according to which the APs serve the ongoing flows. Let $L_{APP-DATA}$ denote the size (in bits) of application-level data per TCP DATA packet. Let $\Theta$ denote the rate (in bits/sec) at which a single isolated AP transfers application-level data to its STAs under long-lived downloads. Then, we have 
\[\quad \quad \quad \quad \quad \Theta = \theta_{singlecell}^{AP-TCP} \times L_{APP-DATA},\] 
where we recall that $\theta_{singlecell}^{AP-TCP}$ denotes the throughput in packets/sec of the AP in a single isolated cell under long-lived downloads. It was shown by \cite{wanet.harsha07WiNet} and \cite{wanet.bruno08TCPeqvSatModel} that $\theta_{singlecell}^{AP-TCP}$ is largely insensitive to the number of STAs in the cell (which is why in Section \ref{subsec:TCP-traffic} we could replace the arbitrary number of STAs by a single saturated STA). Hence, $\Theta$ is also insensitive to the number of STAs in the cell. In our service model, $\Theta$ represents the service rate in bits/sec of a single isolated (processor sharing) queue, which could represent a single isolated cell under short-lived downloads. 

%\vspace{-2mm}

\subsubsection{The Service Model of \cite{wanet.bonald08multicellprocsharing}}
\label{subsubsec:inaccuracy-Bonald-service-model}

%\textbf{The Service Model by Bonald et
%  al. \cite{wanet.bonald08multicellprocsharing}:}

In \cite{wanet.bonald08multicellprocsharing}, different STAs belonging to the same cell might block and be blocked by different subsets of APs and STAs in the other cells, and thus, the users in the same cell could be divided into multiple classes accordingly (see \cite{wanet.bonald08multicellprocsharing} for details). In our network setting, due to the PBD condition (A1), nodes in the same cell have an identical set of interferers, and they belong to the same class. Thus, our network model is a special case of the more general network model in \cite{wanet.bonald08multicellprocsharing}. 

Let $\ind{\cdot}$ denote the indicator function. Let $\varphi_i(\bmath{z})$ denote the rate in bits/sec at which application-level data is transferred by AP-$i$ to its STAs in state $\bmath{z}$, according to the service model in \cite{wanet.bonald08multicellprocsharing}. It turns out that, the service rate $\varphi_i(\bmath{z})$ in bits/sec of the $i^{th}$ processor-sharing queue in state $\bmath{z}$, $z_i >0$, as applied to our network model (of single user class per cell), is given by 
\begin{equation}
\label{eqn:phi-bonald-service-for-single-class-per-cell}
\quad \quad \quad \quad \quad \quad \varphi_i(\bmath{z}) = \frac{\Theta}{1 + \sum_{j \in \mathcal{N}_i} \ind{z_j > 0}}.
\end{equation}

Thus, for our network model, the service model in \cite{wanet.bonald08multicellprocsharing} says that \textit{the service rate of an AP in any state is equal to the service rate of an isolated AP divided by ``1 plus the number of neighboring APs that are non-empty in that state.''} Clearly, the service model in \cite{wanet.bonald08multicellprocsharing} is based on the assumption that, on the average, the single cell capacity is equally shared among the contending neighbors, i.e., among the neighboring APs that have non-empty queues. 

We shall refer to the service model in \cite{wanet.bonald08multicellprocsharing}, restricted to our network model and given by \eqref{eqn:phi-bonald-service-for-single-class-per-cell}, as \textit{Model-1}. In Section \ref{sec:results} we provide an example to show that Model-1, which is based only on the \textit{number} of contending neighbors and does not capture the impact of the overall \textit{topology} of non-empty APs, can lead to inaccurate results.

The fundamental assumption which leads to ``an equal sharing of capacity among contending neighbors'' is the assumption of ``synchronous and slotted evolution'' \cite{wanet.bonald08multicellprocsharing} which is valid if the network consists of a single contention domain, i.e., if $\mathcal{G}$ is \textit{complete} (or, more generally, if $\mathcal{G}$ is a union of complete subgraphs). (Recall that every maximal clique of $\mathcal{G}$ corresponds to a contention domain.) However, as discussed in Section \ref{subsec:saturated-case}, the evolution of activities in networks with multiple overlapping contention domains is \textbf{not} synchronous owing to asynchronous inter-cell blocking. Thus, Model-1 leads to inaccurate results in such networks. 

Figure \ref{fig:threeCellLinear} provides the simplest multi-cell network for which Model-1 leads to inaccurate results (see Example \ref{ex:three-cell-linear} in the next subsubsection). 

%\vspace{-2mm}

\subsubsection{Our Improved Service Model}
\label{subsubsec:our-improved-service-model}

We now propose our improved service model with short-lived downloads as follows. Let 
\[\quad \quad \quad \quad \quad \quad \quad S_{\bmath{z}} := \{i \in \mathcal{N} \; : \; z_i > 0\}\] 
denote the set of cells with non-empty AP queues in state $\bmath{z}$. Let $\mathcal{G}_{\bmath{z}}$ denote the subgraph of $\mathcal{G}$ restricted to $S_{\bmath{z}}$. The service rate $\phi_i(\bmath{z})$ in bits/sec of AP-$i$ in state $\bmath{z}$, $z_i > 0$, in our improved service model is given by 
\begin{equation}
\label{eqn:phi-improved}
%\quad \quad \quad \quad \quad \phi_i(\bmath{z}) = \Theta_i^{app}(\mathcal{G}_{\bmath{z}}) \approx x_i(\mathcal{G}_{\bmath{z}}) \cdot \Theta
\quad \quad \quad \quad \quad \quad \quad \phi_i(\bmath{z}) = x_i(\mathcal{G}_{\bmath{z}}) \cdot \Theta
\end{equation} 
where $x_i(\mathcal{G}_{\bmath{z}})$ can be approximated by the infinite $\rho$ approximation as 
\begin{equation}
\label{eqn:x_i-non-saturated-AP}
\quad \quad \quad \quad \quad \quad \quad x_i(\mathcal{G}_{\bmath{z}}) \approx \frac{\eta_i\big(\mathcal{G}_{\bmath{z}}\big)}{\eta\big(\mathcal{G}_{\bmath{z}}\big)}.
\end{equation}

Thus, in any state $\bmath{z}$, we (i) obtain the subset $S_{\bmath{z}}$ of cells with non-empty AP queues, (ii) obtain the restriction $\mathcal{G}_{\bmath{z}}$ of the contention graph $\mathcal{G}$ by restricting $\mathcal{G}$ to the cells in $S_{\bmath{z}}$, (iii) obtain the normalized throughputs of the APs in $S_{\bmath{z}}$ w.r.t.~the contention graph $\mathcal{G}_{\bmath{z}}$ by invoking the infinite $\rho$ approximation, and (iv) obtain the state-dependent service rates in state $\bmath{z}$ by taking the product of normalized throughputs and the application-level single cell throughput  $\Theta$. 

We shall refer to our service model, given by \eqref{eqn:phi-improved}, as \textit{Model-2}, and demonstrate its accuracy in Section \ref{sec:results}. We emphasize that \textit{Model-2 reduces to Model-1 for networks with a single contention domain}. However, for networks with multiple overlapping contention domains, Model-2 can capture the impact of the overall topology of the network which Model-1 cannot.

\begin{example}
\label{ex:three-cell-linear}
Consider Figure \ref{fig:threeCellLinear} and suppose that the APs in all the three cells have non-empty queues, i.e., $z_1, z_2, z_3 > 0$. According to Model-1, the ratio of their service rates should be $\frac{1}{2}:\frac{1}{3}:\frac{1}{2}$. However, according to Model-2, the ratio of their service rates should be $1:0:1$. In NS-2 simulations, if we saturate the AP queues so as to keep them non-empty all the time, the ratio of throughputs is observed to be $0.96:0.04:0.96$ which supports Model-2 rather than Model-1. Thus, as shown in Section \ref{sec:results}, Model-1 provides inaccurate results, but Model-2 is quite accurate. \hfill \qed
\end{example}

%\vspace{-2mm}

\subsection{Approximate Delay Analysis}
\label{subsubsec:delay-analysis}

Let $\hat{x}_i$ denote the long-term average normalized throughput of AP-$i$, i.e., $\hat{x}_i$ denotes the long-term average fraction of time for which the nodes in Cell-$i$ are not blocked. The \textit{``effective service rate''} $\hat{\phi}_i$ in bits/sec of the $i^{th}$ processor-sharing queue is then given by $\hat{\phi}_i = \hat{x}_i \Theta$. 

We decouple the queues by assuming that each queue $i$ evolves as an independent processor-sharing queue with arrival rate $\nu_i$ flows/sec, i.i.d.~exponentially distributed flow sizes with mean $\EXP{V}$ bits and service rate $\hat{\phi}_i$ bits/sec. The probability that queue $i$ is empty (resp.~non-empty) is then given by $\max\{0,1-\frac{\nu_i \EXP{V}}{\hat{x}_i \Theta}\}$ (resp.~$\min\{1,\frac{\nu_i \EXP{V}}{\hat{x}_i \Theta}\}$).

We model the inter-dependence among the queues by requiring that, $\forall i \in \mathcal{N}$, the $\hat{x}_i$'s satisfy Equation \eqref{eqn:fixed-point-x_i} (placed at the top of the next page). Equation \eqref{eqn:fixed-point-x_i} is explained as follows. We consider the queues, other than queue $i$, and obtain the probability $p(S)$ that the subset $S \subseteq \mathcal{N} \setminus \{i\}$ of queues is non-empty, assuming the queues to be independent given the effective service rates $\hat{\phi}_j$'s. We then multiply $p(S)$ with the normalized throughput of AP-$i$ w.r.t.~the graph $\mathcal{G}[S \cup \{i\}]$, where $\mathcal{G}[S \cup \{i\}]$ denotes the restriction of $\mathcal{G}$ to the subset $S \cup \{i\}$ of vertices. Finally, we sum over all possible subsets $S \subseteq \mathcal{N} \setminus \{i\}$ and obtain \eqref{eqn:fixed-point-x_i}.

\begin{figure*}[tbh]
\begin{footnotesize}
\begin{equation}
\label{eqn:fixed-point-x_i}
\quad \quad \quad \quad \quad \hat{x}_i = \displaystyle \sum_{S \subseteq \mathcal{N} \setminus
  \{i\}} \left( \prod_{j \in S} \min \left\{1,\frac{\nu_j \EXP{V}}{\hat{x}_j
  \Theta} \right\} \right) \times \left( \prod_{k \in \mathcal{N} \setminus
  (S \cup \{i\})} \max \left\{0,1-\frac{\nu_k \EXP{V}}{\hat{x}_k
  \Theta} \right\}\right) \times \left( \frac{\eta_i\big(\mathcal{G}[S \cup
  \{i\}]\big)}{\eta\big(\mathcal{G}[S \cup \{i\}]\big)} \right)
%\; \;\; \; \; \; \; \; \; \; \; \; \;
\end{equation}
\end{footnotesize}
\rule{173mm}{.3mm}
%\vspace*{-7mm}
\end{figure*}

The set of equations given by Equation \eqref{eqn:fixed-point-x_i} constitute a multi-dimensional fixed-point equation which can be solved to obtain the $\hat{x}_i$'s. Once the $\hat{x}_i$'s are obtained, we approximate the \textit{mean flow transfer delay} $\EXP{D}_i$ in Cell-$i$ by 
\[\quad \quad \quad \quad \quad \quad \quad \EXP{D}_i = \frac{\frac{\EXP{V}}{\hat{x}_i \Theta}}{1 - \frac{\nu_i \EXP{V}}{\hat{x}_i \Theta}}\] 
which resembles the standard expression for mean flow transfer delay in a processor-sharing queue with mean arrival rate $\nu_i$ flows/sec, mean flow size $\EXP{V}$ bits, and service rate $\hat{x}_i \Theta$ bits/sec. Results obtained from our approximate delay analysis are provided in Section \ref{subsec:results-TCP-short}.

\section{Numerical and Simulation Results}
\label{sec:results}

In this section we validate our analytical model by comparing with the results obtained from NS-2 simulations \cite{wanet.ns2}. We created the example topologies in Figures \ref{fig:twoCell}-\ref{fig:sevenCellArbitrary}. We chose the cell radii and the carrier sensing range to ensure that the PBD condition is satisfied. Under the PBD condition, the collision model in NS-2 is exactly as given by Assumptions (A2)-(A3). The channel was configured to be error-free. Nodes were randomly placed within the cells. The AP buffer size was set large enough to avoid buffer losses. The delayed ACK mechanism was disabled. Each case was simulated 20 times, each run for 200 sec of ``network time''. (The short-file download case was simulated for 1000 flows per cell per run.) We report the results for ``Basic Access''. We took 11 Mbps data and control rates and data payload size of 1000 Bytes. We report only the mean values of simulation results. In almost all cases, the 99\% confidence intervals about the mean values were observed to be within 5-10\% of the mean values.

%We disabled EIFS deferrals by modifying the NS-2 simulator such that medium access always begins/resumes after a DIFS deferral. 

%The saturated cases were simulated with high rate CBR over UDP
%connections. For the TCP cases, we created one TCP download connection
%per STA. Each TCP connection was fed by an FTP source with the TCP
%source agents attached directly to the AP to emulate a local server.

%The function ``\textit{fsolve()}'' of MATLAB was used for solving the fixed point equations. 

The analytical values of single cell throughputs per-node for the saturated case were obtained by applying the model of \cite{wanet.kumar_etal07new_insights}. The analytical values of single cell throughputs of the APs for the TCP case were obtained by applying the model of \cite{wanet.bruno08TCPeqvSatModel}. The analytical single cell throughputs per-node (resp.~of the AP) multiplied with the $x_i$'s obtained from our multi-cell model provide the analytical throughputs per-node (resp.~of the AP) in a multi-cell network (see \eqref{eqn:Theta-multicell}).

%Results for the saturated case are summarized in Section
%\ref{subsec:results-saturated}, that for long-lived downloads are
%summarized in Section \ref{subsec:results-TCP-long}, and that for
%short-lived downloads are summarized in Section
%\ref{subsec:results-TCP-short}. Results for two network scenarios
%where the PBD condition does not hold are provided in Section
%\ref{subsec:when-PBD-does-not-hold}. More numerical and simulation
%results in support of our analytical model can be found in
%\cite{wanet.manoj-anurag10multicellTechReportWithShortTCP}.

%\vspace{-2mm}

\subsection{Saturated Nodes with UDP Traffic}
\label{subsec:results-saturated}

Figures \ref{fig:SevenCellArbitraryGamma} and \ref{fig:SevenCellArbitraryTheta} compare the collision probability $\gamma$ and the throughput per node $\theta$, respectively, for the example network in Figure \ref{fig:sevenCellArbitrary} with saturated nodes and UDP traffic. For comparison, in Figures \ref{fig:SevenCellArbitraryGamma} and \ref{fig:SevenCellArbitraryTheta} we also show the corresponding single cell results, i.e., the results one would expect had the seven cells been mutually independent. Referring to Figures \ref{fig:SevenCellArbitraryGamma} and \ref{fig:SevenCellArbitraryTheta}, we make the following observations:

\begin{figure}[tb]
\centering
\begin{minipage}{8.25cm}
\begin{center}
\includegraphics[scale=0.95]{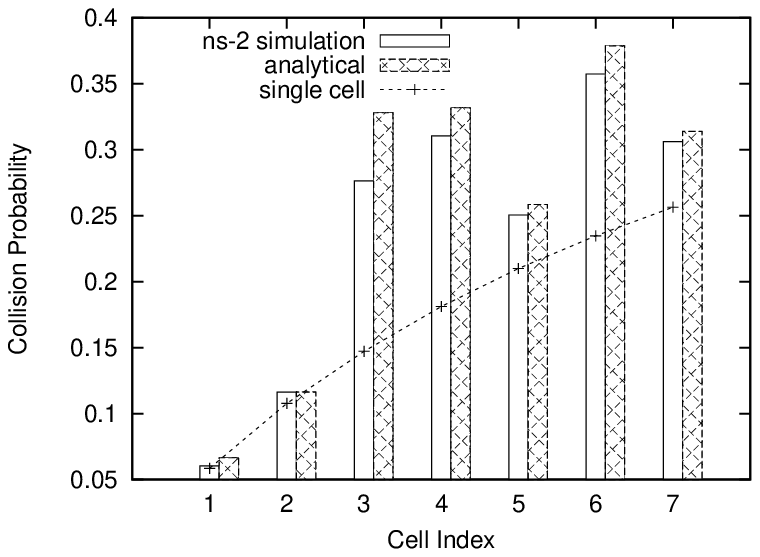}
\caption{Collision probability $\gamma$ for the scenario in Figure \ref{fig:sevenCellArbitrary} with UDP traffic when Cell-$i$, $1 \leq i \leq 7$, contains $n_i = i+1$ saturated nodes.\label{fig:SevenCellArbitraryGamma}}
\vspace{2mm}
\end{center}
\end{minipage}
\hfill
\begin{minipage}{8.25cm}
\begin{center}
\includegraphics[scale=0.95]{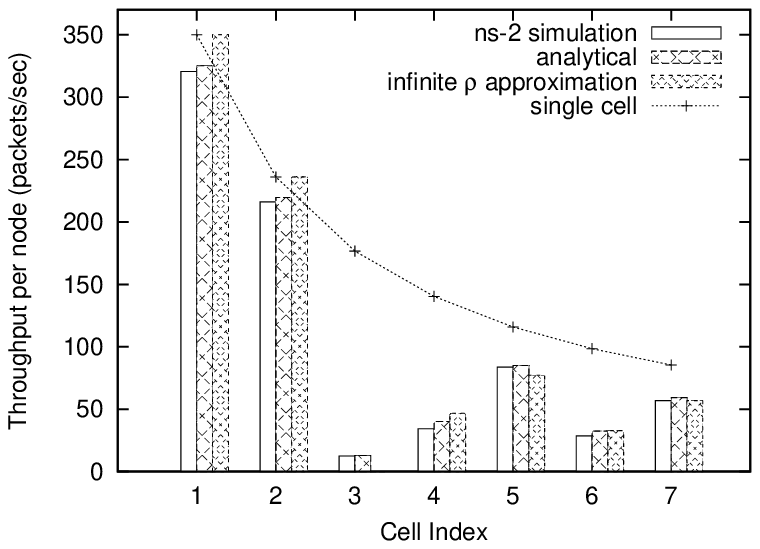}
\caption{Per-node throughput $\theta$ for the scenario in Figure \ref{fig:sevenCellArbitrary} with UDP traffic when Cell-$i$, $1 \leq i \leq 7$, contains $n_i = i+1$ saturated nodes.\label{fig:SevenCellArbitraryTheta}}
\end{center}
\end{minipage}
%\vspace{-2mm}
\end{figure}

\begin{itemize}

\item [O$_1$] Our fixed-point analysis provides quite accurate predictions of collision probabilities and throughputs (less than 10\% error in most cases). The slight over-estimation of throughputs is due to the approximation in \eqref{eqn:Theta-multicell}. The infinite $\rho$ approximation (which does not require solving a fixed-point equation) provides a simple and efficient method to quickly compute the throughputs fairly accurately. Nevertheless, our fixed point analysis provides more accurate predictions for throughputs and also provides accurate predictions for collision probabilities which cannot be provided by the infinite $\rho$ approximation. Had we not accounted for the inter-cell collisions in the second stage, our analytical collision probabilities would have been equal to the corresponding single cell collision probabilities (see Figure \ref{fig:SevenCellArbitraryGamma}). 

\item [O$_2$] The throughput of a cell cannot be accurately determined based only on the \textit{number} of neighboring cells. In Figure \ref{fig:sevenCellArbitrary}, Cell-3 and Cell-4 each have three neighbors but their per node throughputs $\theta$ are quite different (see Figure \ref{fig:SevenCellArbitraryTheta}). In particular, $\theta_4 > \theta_3$ even though $n_3 = 4 < n_4 = 5$. This is due to Cell-7 which blocks Cell-6 for certain fraction of time during which Cell-4 gets opportunity to transmit whereas Cell-1 and Cell-2 are almost never blocked and Cell-3 is almost always blocked due to Cell-1 and Cell-2.  Thus, \textit{topology of the entire network plays the key role}. 

\end{itemize}

\subsection{Long-lived Downloads}
\label{subsec:results-TCP-long}

\begin{figure}[tb]
\centering
\begin{minipage}{8.25cm}
\begin{center}
\includegraphics[scale=0.95]{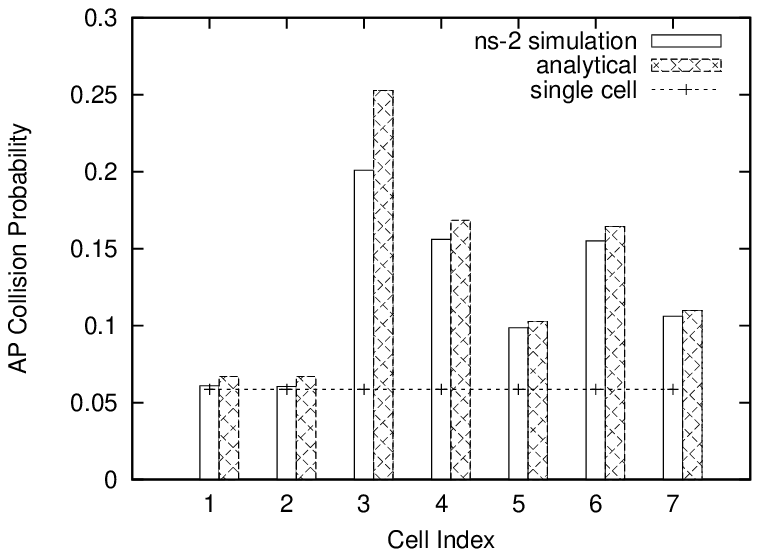}
\caption{Collision probability $\gamma$ for the example
  scenario in Figure \ref{fig:sevenCellArbitrary} when each cell
  contains an AP and $n = 5$ STAs.  STAs are downloading long files
  through their respective APs using TCP
  connections.\label{fig:SevenCellArbitraryGammaTCP}}
  \vspace{2mm}
\end{center}
\end{minipage}
\hfill
\begin{minipage}{8.25cm}
\begin{center}
\includegraphics[scale=0.95]{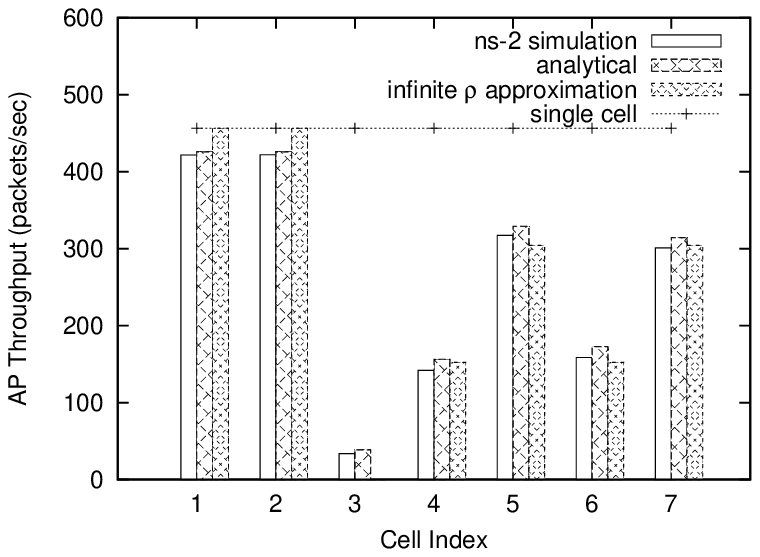}
\caption{Throughput per node $\theta$ for the example
  scenario in Figure \ref{fig:sevenCellArbitrary} when each cell
  contains an AP and $n = 5$ STAs.  STAs are downloading long files
  through their respective APs using TCP
  connections.\label{fig:SevenCellArbitraryThetaTCP}}
\end{center}
\end{minipage}
%\vspace{-2mm}
\end{figure}

Figures \ref{fig:SevenCellArbitraryGammaTCP} and \ref{fig:SevenCellArbitraryThetaTCP} compare the collision probability $\gamma$ and the throughput of the AP, respectively, for the example network in Figure \ref{fig:sevenCellArbitrary}. Our simulation results in this case correspond to $n = 5$ STAs in each cell. Note, however, that \textit{the analytical values in Figures \ref{fig:SevenCellArbitraryGammaTCP} and \ref{fig:SevenCellArbitraryThetaTCP} were obtained by taking $n_i = 2$ saturated nodes in every cell}. Referring to Figures \ref{fig:SevenCellArbitraryGammaTCP} and \ref{fig:SevenCellArbitraryThetaTCP}, we conclude that the foregoing observations (O$_1$) and (O$_2$) for the saturated case carry over to TCP-controlled long file downloads as well. 

Referring to Figure \ref{fig:SevenCellArbitraryThetaTCP}, we can compare the state-dependent service rates as per Model-1 and Model-2. Model-1, the service model of \cite{wanet.bonald08multicellprocsharing}, would predict the throughput ratios as $\frac{1}{2} : \frac{1}{2} : \frac{1}{4} : \frac{1}{4} : \frac{1}{2} : \frac{1}{3} : \frac{1}{2}$. The observed throughput ratios in Figure \ref{fig:SevenCellArbitraryThetaTCP} are much closer to Model-2's prediction given by $1 : 1 : 0 : \frac{1}{3} : \frac{2}{3} : \frac{1}{3} : \frac{2}{3}$.

\subsection{Relaxing the PBD Condition in Simulations}
\label{subsec:when-PBD-does-not-hold}

%We demonstrated that our analytical model is able to predict the collision probabilities and the cell throughputs quite accurately if the PBD condition holds. 

\begin{figure}[tb]
\centering
\hspace{-4mm}
\begin{minipage}{4.5cm}
\begin{center}
\includegraphics[scale=0.35]{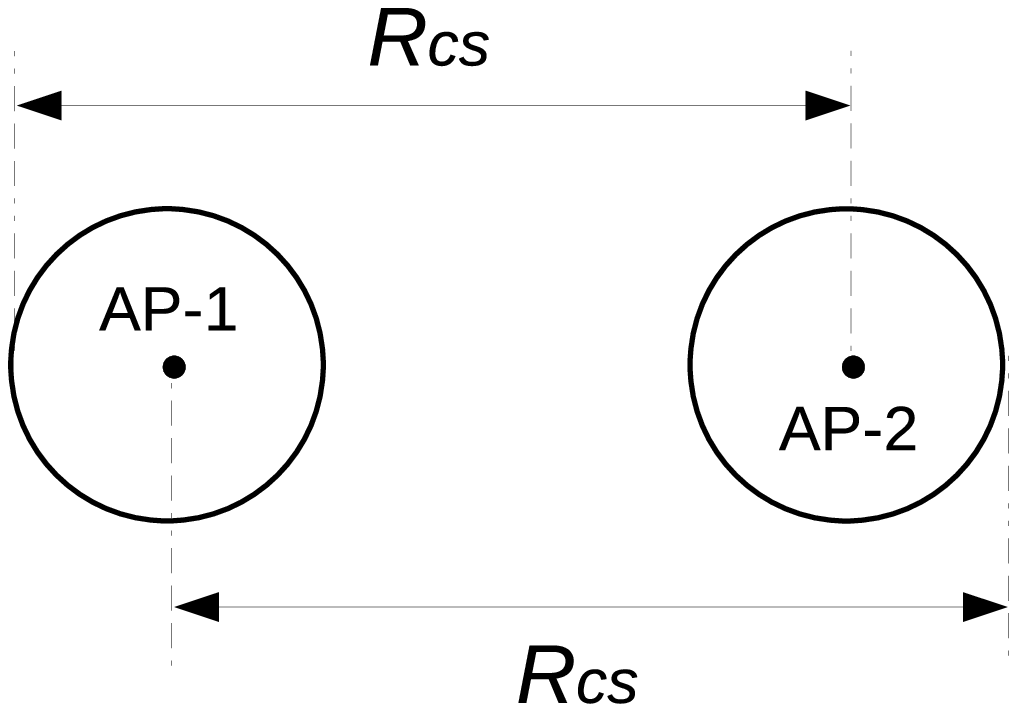}
\caption{Relaxation (R1) of the PBD condition: APs can sense all the
  nodes in the other cell, but $\approx 50\%$ STA(s) can sense only a
  subset of STAs in the other cell. \label{fig:relaxation1}}
\end{center}
\end{minipage}
%\hfill
\hspace{1mm}
\begin{minipage}{3.2cm}
\begin{center}
\includegraphics[scale=0.35]{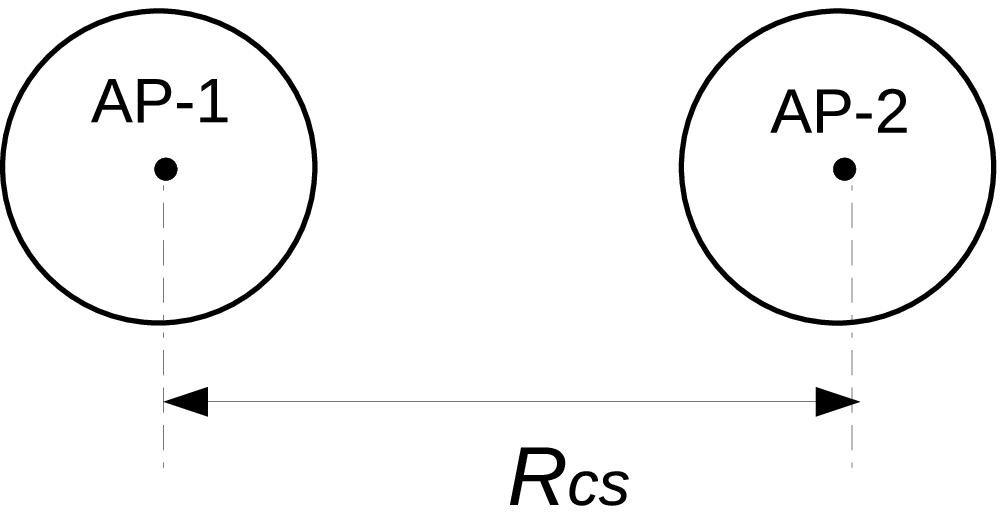}
\caption{Relaxation (R2) of the PBD condition: APs can sense each
  other, but the APs cannot sense $\approx 50\%$ STA(s) in the other
  cell. \label{fig:relaxation2}}
\end{center}
\end{minipage}
%\vspace{-3mm}
\end{figure}

In this subsection we report simulation results for two network scenarios where the PBD condition does not hold, i.e., when two dependent cells are not necessarily completely dependent (recall Definition \ref{defn:dependence}) such that only a subset of nodes in one cell are dependent w.r.t.~a subset of nodes in the other cell. We consider TCP-controlled long-lived downloads where the APs transmit for a larger fraction of time than the STAs and examine two ``relaxations'' of the PBD condition. The two relaxations define the nature of dependence between two dependent cells as follows:

\begin{itemize}

\item [R1] For any two dependent cells, the APs can sense all the nodes in the other cell, but $\approx 50\%$ STA(s) can sense only a (proper) subset of STAs in the other cell. Figure \ref{fig:relaxation1} depicts this dependence.

\item [R2] For any two dependent cells, the APs can sense each other, but the APs cannot sense $\approx 50\%$ STA(s) in the other cell. Figure \ref{fig:relaxation2} depicts this dependence.

\end{itemize}

Table \ref{table:results-when-PBD-does-not-hold} summarizes the simulation results for the seven arbitrarily placed cells in Figure \ref{fig:sevenCellArbitrary} where \textbf{the edges in the contention graph now represent dependence according to either R1 or R2} and not complete dependence. Each cell contains 1 AP and $n=10$ STAs. For comparison, we also provide the analytical results (which are obtained by our fixed-point analysis) under the PBD condition. Referring Table \ref{table:results-when-PBD-does-not-hold} we make the following observations:
\begin{itemize}

\item [O$_3$] Our analytical model roughly captures the throughput
  distribution even when the PBD condition does not hold.  In
  particular, it predicts that Cell-3 is blocked for a large fraction
  of time as compared to the remaining cells.  Note that, a heuristic
  model based only on the number of interfering APs would have
  predicted equal throughputs for Cell-3 and Cell-4 which is not the
  case.
\end{itemize}

\begin{small}
\begin{table}[tb]
\begin{center}
\caption{\label{table:results-when-PBD-does-not-hold} Results with (R1) and (R2) dependence: All throughputs are in packets/sec.}
%\vspace{4mm}
\begin{footnotesize}
\begin{tabular}{||c|c|c|c|c|c|c||}
\hline
& \multicolumn{2}{c|}{} & \multicolumn{2}{c|}{} & \multicolumn{2}{c||}{} \\
& \multicolumn{2}{c|}{R1 Dependence} & \multicolumn{2}{c|}{R2 Dependence} & \multicolumn{2}{c||}{Analytical} \\
\cline{2-7} & & & & & & \\
Cell & $\gamma_{sim}^{AP}$ & $\theta_{sim}^{AP}$ & $\gamma_{sim}^{AP}$ & $\theta_{sim}^{AP}$ & $\gamma_{ana}^{AP}$ & $\theta_{ana}^{AP}$ \\
index & & & & & & \\
%index & & (pkts/sec) & & (pkts/sec) & & (pkts/sec) \\
\hline
\hline
1 & 0.061 & 424.85 & 0.109 & 359.57 & 0.067 & 425.83 \\
\hline
2 & 0.064 & 422.07 & 0.096 & 366.68 & 0.067 & 425.83 \\
\hline
3 & 0.317 & 25.95 & 0.319 & 96.87 & 0.253 & 38.50 \\
\hline
4 & 0.266 & 110.19 & 0.269 & 156.39 & 0.169 & 156.41 \\
\hline
5 & 0.102 & 321.41 & 0.101 & 378.74 & 0.103 & 329.06 \\
\hline
6 & 0.174 & 164.33 & 0.197 & 170.16 & 0.164 & 172.64 \\
\hline
7 & 0.124 & 285.47 & 0.148 & 296.06 & 0.110 & 314.10 \\
\hline
\end{tabular}
\end{footnotesize}
\end{center}
%\vspace{-4mm}
\end{table}
\end{small}

\subsection{Short-lived Downloads}
\label{subsec:results-TCP-short}

To compare Model-1 and Model-2, we developed a customized event-driven queueing simulator which can simulate networks of single-class processor-sharing queues with Poisson flow arrivals, i.i.d.~exponentially distributed flow sizes and the two service models, namely, Model-1 and Model-2. Results from the model-based flow-level queueing simulations are compared against the results obtained with the NS-2 simulator \cite{wanet.ns2} which performs protocol-based packet-level detailed simulations with $n = 5$ STAs per cell. We also compare the above simulation results with our approximate delay analysis.

\begin{figure}[tb]
  \centering
  \begin{minipage}{8.25cm}
  \begin{center}
    \includegraphics[scale=0.8]{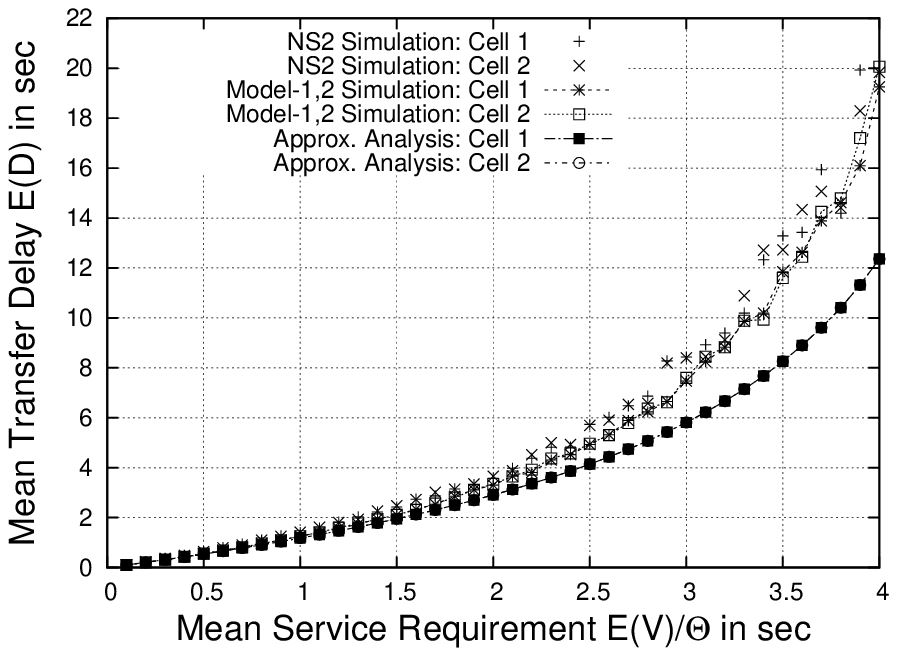}
     \caption{Comparing the mean flow transfer delays for the two-cell
       network in Figure \ref{fig:twoCell} with $\nu_1 = \nu_2 = 0.1
       \; sec^{-1}$. \label{fig:shortTCPDelayVsEVTwoCellNuPoint1Point1Few}}
     \vspace{3mm}
  \end{center}
  \end{minipage}
%  \hfill
  \begin{minipage}{8.25cm}
  \begin{center}
    \includegraphics[scale=0.8]{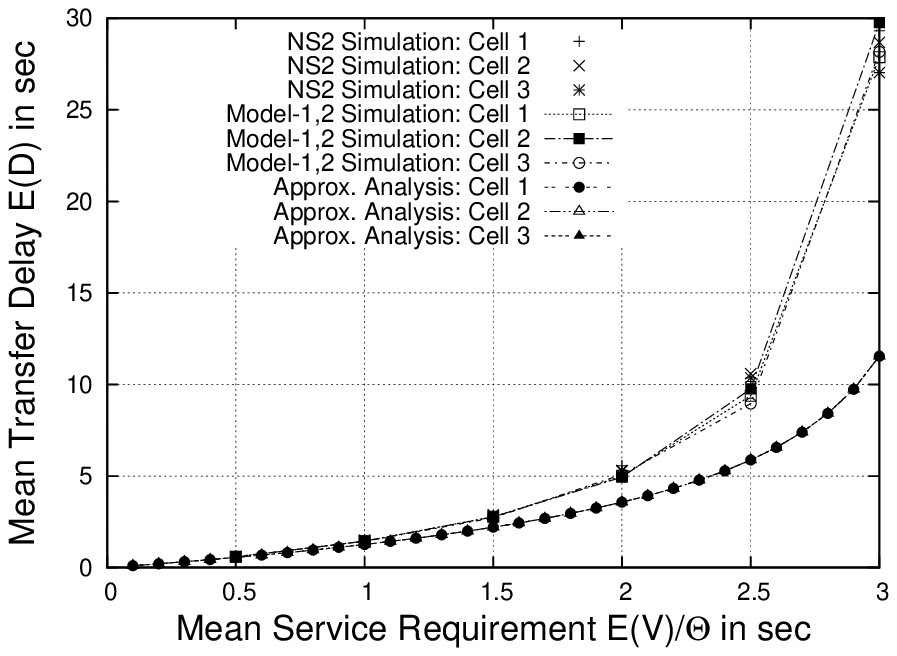}
     \caption{Comparing the mean flow transfer delays for the
       three-cell network in Figure \ref{fig:threeCellSingleCD} with
       $\nu_1 = \nu_2 = \nu_3 = 0.1 \; sec^{-1}$. \label{fig:shortTCPDelayVsEVNuPoint1ThreeCellSingleCD}}
     \vspace{3mm}
  \end{center}
  \end{minipage}
  \begin{minipage}{8.25cm}
    \begin{center}
      \includegraphics[scale=0.8]{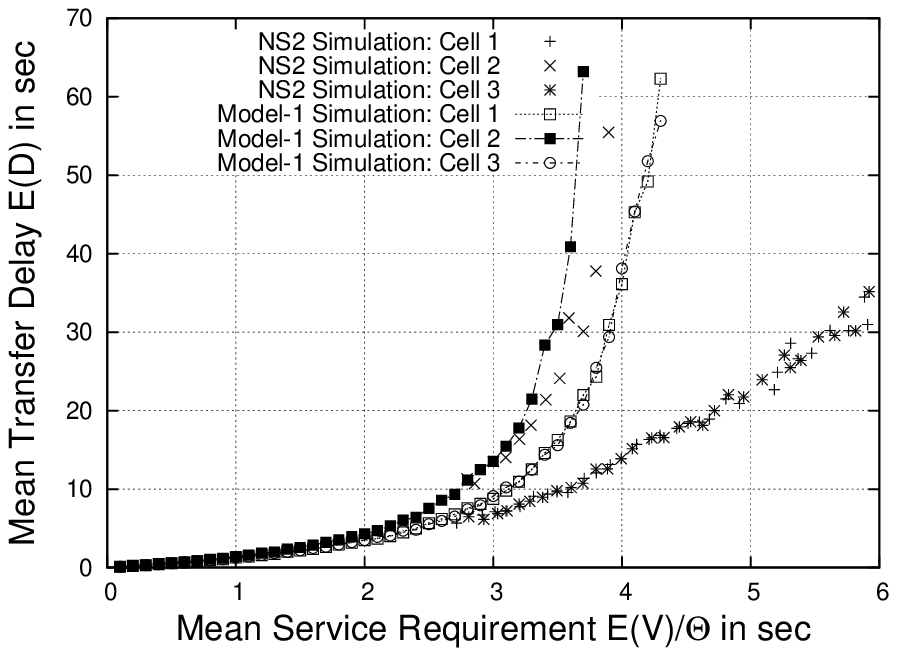}
     \caption{Comparing the mean flow transfer delays from
       NS-2 simulations with the delays corresponding to Model-1 for
       the three-cell network in Figure \ref{fig:threeCellLinear}
       with $\nu_1 = \nu_2 = \nu_3 = 0.1 \;
       sec^{-1}$. \label{fig:shortTCPDelayVsEVNuPoint1ThreeCellLinearModel1}}
%     \vspace{5mm}
    \end{center}
  \end{minipage}
%  \vspace{-2mm}
\end{figure}

\subsubsection{Single Contention Domain}
\label{subsubsec:single-contention-domain}

Figure \ref{fig:shortTCPDelayVsEVTwoCellNuPoint1Point1Few} compares
the mean flow transfer delays obtained from (i) NS-2 simulations, (ii)
our customized queueing simulations according to Model-1 and Model-2, and
(iii) our approximate delay analysis, for the two-cell network in
Figure \ref{fig:twoCell} with $\nu_1 = \nu_2 = 0.1 \;
sec^{-1}$. Figure \ref{fig:shortTCPDelayVsEVNuPoint1ThreeCellSingleCD}
shows a similar comparison for the three-cell network in Figure
\ref{fig:threeCellSingleCD}. The $x$-axis depicts the \textit{mean
  service requirement}, $\frac{\EXP{V}}{\Theta}$, of flows in
seconds. For example, $\frac{\EXP{V}}{\Theta} =  3 \; sec$,
corresponds to exponentially distributed flow sizes of mean $\EXP{V}$
bits where the value of $\EXP{V}$ is such that, a server which works
at a constant rate of $\Theta$ bits/sec has to dedicate $3 \; sec$ to
completely serve a flow. Since the two-cell (resp.~three-cell) network
in Figure \ref{fig:twoCell} (resp.~Figure \ref{fig:threeCellSingleCD})
consists of a single contention domain, Model-1 and Model-2 are
identical for these cases. We observe that: 
\begin{itemize}

\item [O$_4$] A close match between our customized queueing simulations 
and NS-2 simulations validates Model-1 (also Model-2) for networks with 
a single contention domain.  

\end{itemize}

\begin{figure}[tb]
  \centering
  \begin{minipage}{8.25cm}
    \begin{center}
      \includegraphics[scale=0.85]{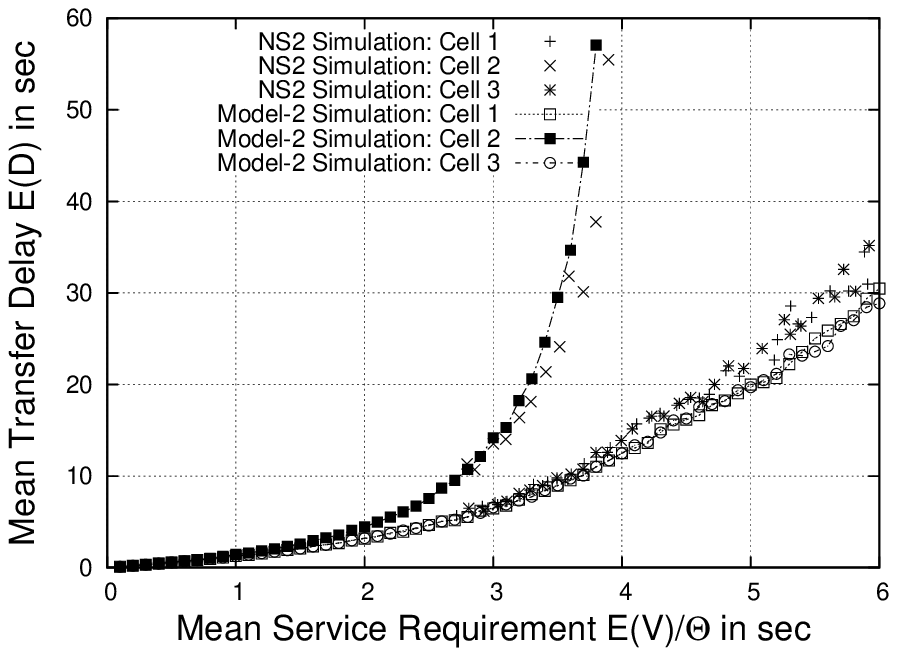}
     \caption{Comparing the mean flow transfer delays from
       NS-2 simulations with the delays corresponding to Model-2 for
       the three-cell network in Figure \ref{fig:threeCellLinear}
       with $\nu_1 = \nu_2 = \nu_3 = 0.1 \;
       sec^{-1}$. \label{fig:shortTCPDelayVsEVNuPoint1ThreeCellLinearModel2}}
     \vspace{2mm}
    \end{center}
  \end{minipage}
%  \hfill
  \begin{minipage}{8.25cm}
    \begin{center}
      \includegraphics[scale=0.85]{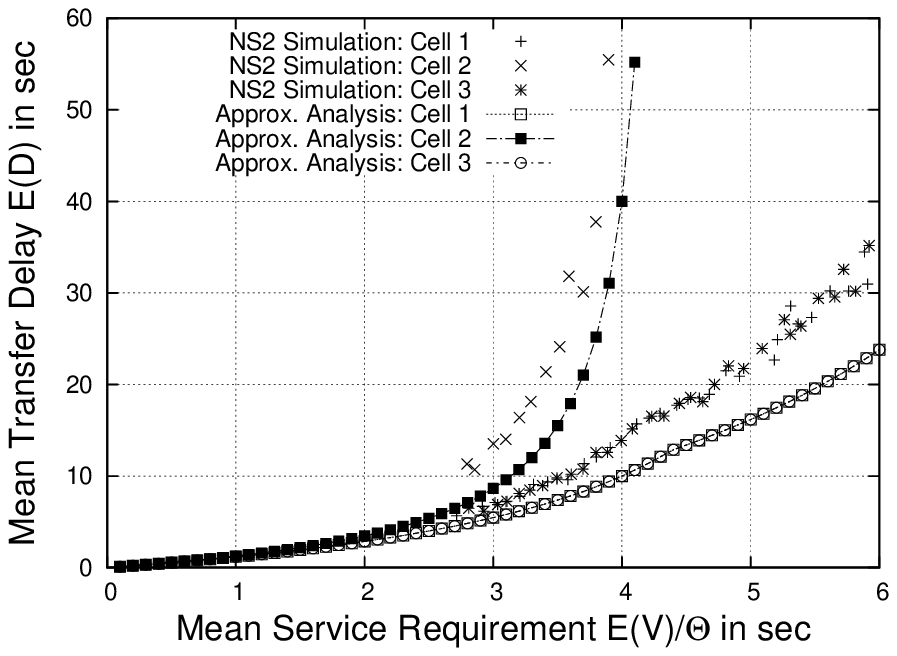}
     \caption{Comparing the mean flow transfer delays from
       NS-2 simulations with the delays from our approximate delay
       analysis for the three-cell network in Figure
       \ref{fig:threeCellLinear} with $\nu_1 = \nu_2 = \nu_3 = 0.1
       \; sec^{-1}$. Note that the points corresponding to our
       approximate delay analysis for Cell-1 and Cell-3 are
       overlapping. \label{fig:shortTCPDelayVsEVNuPoint1ThreeCellLinearApproxAnal}}
    \end{center}
  \end{minipage}
%\vspace{-2mm}
\end{figure}

\subsubsection{Multiple Overlapping Contention Domains}
\label{subsubsec:multiple-contention-domain}

Figures \ref{fig:shortTCPDelayVsEVNuPoint1ThreeCellLinearModel1},
\ref{fig:shortTCPDelayVsEVNuPoint1ThreeCellLinearModel2}, and
\ref{fig:shortTCPDelayVsEVNuPoint1ThreeCellLinearApproxAnal} compare
the mean flow transfer delays from NS-2 simulations with the mean flow
transfer delays corresponding to Model-1, Model-2, and our approximate
delay analysis, respectively, for the three-cell network in Figure
\ref{fig:threeCellLinear} with $\nu_i = 0.1 \; sec^{-1}$. We observe 
that: 
\begin{itemize}

\item [O$_{5}$] Model-1 leads to inaccurate results, since it 
predicts delays for Cell-1 and Cell-3 that are much higher than that 
obtained from NS-2 simulations (see Figure 
\ref{fig:shortTCPDelayVsEVNuPoint1ThreeCellLinearModel1}). Apart from 
the mismatch, Model-1 does not capture the way the network behaves with variation in load. However, 
Model-2 matches with NS-2 simulations extremely well (see Figure
\ref{fig:shortTCPDelayVsEVNuPoint1ThreeCellLinearModel2}). 

\item [O$_{6}$] Our approximate delay analysis provides good 
predictions for the mean transfer delays (see Figure 
\ref{fig:shortTCPDelayVsEVNuPoint1ThreeCellLinearApproxAnal}). 

\end{itemize}

%\section{Application to Channel Assignment}
%\label{sec:channel-assignment}
%
%The insights provided by our cell-level model could be applied to
%design a fast and distributed channel assignment algorithm called mISA
%\cite{wanet.manoj-anuragRAWNET09multicell}. In
%\cite{wanet.manoj-anuragRAWNET09multicell}, we show that the channel
%assignments provided by mISA are \textit{Nash equilibria in pure
%  strategies} for the objective of maximizing the normalized network
%throughput. Furthermore, mISA provides a channel assignment in only as
%many iterations as there are channels. In
%\cite{wanet.manoj-anuragRAWNET09multicell}, we also show that, mISA
%can be implemented in a completely distributed manner without any
%message passing provided that the APs use identical set of
%non-overlapping channels. We do not provide the details here due to
%space constraints. Interested readers may refer to
%\cite{wanet.manoj-anuragRAWNET09multicell}.

\section{Conclusion}
\label{sec:conclusion}

In this paper, we identified a Pairwise Binary Dependence (PBD) condition which enables scalable cell-level modeling of multi-cell infrastructure WLANs. Under a unified framework, we developed accurate analytical models for multi-cell infrastructure WLANs with saturated nodes, and for TCP-controlled long- and short-file downloads. An important property of CSMA, namely, ``maximization of the network throughput in a distributed manner'' has been established in the recent work by \cite{wanet.durvy09selfOrganization} in the context of an infinite linear chain of saturated nodes. We established it for multi-cell infrastructure WLANs with arbitrary cell topologies under the assumption of the PBD condition. We also provided analytical and simulation results in support of the so-called ``infinite $\rho$ approximation'' which exhibit the throughput maximization property of CSMA. We characterized the service process in multi-cell infrastructure WLANs with TCP-controlled short-lived downloads. Such characterization is essential for accurately computing the mean flow transfer delays. We improved the service model of \cite{wanet.bonald08multicellprocsharing} by incorporating the impact of the \textit{topology} of the entire network and demonstrated its accuracy. By applying an ``effective service rate approximation'' technique we also obtained good approximations for the mean flow transfer delay in each cell.

\appendix

\subsection{Appendix: Derivation of Equation \eqref{eqn:gamma_i-multicell}} 
\label{app:derivation-eqn-gamma_i}

Consider a tagged node in Cell-$i$. Let

\vspace{1mm}

$A_i(T)$ : \parbox[t]{6cm}{number of transmission attempts made by the tagged node up to time $T$}

\vspace{1mm}

$C_i(T)$ : \parbox[t]{6cm}{number of collisions as seen by the tagged node up to time $T$} 

\vspace{1mm}

Then, the (conditional) collision probability $\gamma_i$ of a tagged node in Cell-$i$, i.e., the probability that a transmission attempt in Cell-$i$ collides, is given by
\begin{equation}
\label{eqn:gamma-definition}
\quad \quad \quad \quad \quad \quad \gamma_i := \displaystyle \lim_{T \longrightarrow \infty} \frac{C_i(T)}{A_i(T)}.
\end{equation}

%\begin{equation}
%\label{eqn:beta-definition}
%\beta_i := \displaystyle \lim_{T \rightarrow \infty}
%\frac{A_i(T)}{B_i(T)}. 
%\end{equation}

%\vspace{-2mm}

Consider any state $\mathcal{A}$ such that $i \in \mathcal{U}_{\mathcal{A}}$. Let 

\vspace{1mm}

$A_i^{\mathcal{A}}(T)$ : \parbox[t] {6cm} {number of transmission attempts made by the tagged node in State-$\mathcal{A}$, up to time $T$} 

\vspace{1mm}

$C_i^{\mathcal{A}}(T)$ : \parbox[t] {6cm} {number of collisions as seen by the tagged node, up to time $T$, given that the transmission attempts were made in State-${\mathcal{A}}$}

\vspace{1mm}

$B_i^{\mathcal{A}}(T)$ : \parbox[t] {6cm} {number of backoff slots elapsed in the local medium of Cell-$i$ in State-${\mathcal{A}}$, up to time $T$}

\vspace{1mm}

Since the CTMC $\{\mathcal{A}(t), t \geq 0\}$ is stationary and
ergodic, the (long-term) fraction of time for which the network
remains in State-$\mathcal{A}$ is equal to the stationary probability
$\pi(\mathcal{A})$. Consider now a sufficiently long time $T$. Then,
the time spent in State-$\mathcal{A}$ up to time $T$ is approximately
equal to $\pi(\mathcal{A})T$, and $B_i^{\mathcal{A}}(T) \approx
\displaystyle \frac{\pi(\mathcal{A}) T}{\sigma}$ where recall that
$\sigma$ is the duration of a backoff slot. Thus, we have 
\begin{equation}
\label{eqn:number-of-backoff-slots}
\quad \quad \quad \quad \quad \quad \displaystyle \lim_{T \longrightarrow \infty} \frac{B_i^{\mathcal{A}}(T)}{T} = \frac{\pi(\mathcal{A})}{\sigma}. 
\end{equation}

Since, the nodes in Cell-$i$ attempt with probability $\beta_i$ in
every backoff slot irrespective of whether the other cells can attempt
or not, we have $A_i^{\mathcal{A}}(T) \approx \beta_i
B_i^{\mathcal{A}}(T)$. Thus, 
\begin{equation}
\label{eqn:beta-definition}
\quad \quad \quad \quad \quad \quad \displaystyle \lim_{T \longrightarrow \infty} \frac{A_i^{\mathcal{A}}(T)}{B_i^{\mathcal{A}}(T)} = \beta_i. 
\end{equation}

%Equation \ref{eqn:gamma_i-multicell} can now be derived as follows. 
By definition (see \eqref{eqn:gamma-definition}), 
%\begin{figure*}
%\begin{center}
%%\vspace{-5mm}
%\begin{small}
%\begin{minipage}{10cm}
\begin{eqnarray}
\label{eqn:gamma_i-multicell-derivation}
\gamma_i &=& \displaystyle \lim_{T \longrightarrow \infty} \frac{C_i(T)}{A_i(T)} = \displaystyle \frac{\displaystyle \lim_{T \longrightarrow \infty} \frac{1}{T} \; C_i(T)}{\displaystyle \lim_{T \longrightarrow \infty} \frac{1}{T} \; A_i(T)} \nonumber \\ 
&=& \frac{\displaystyle \lim_{T \longrightarrow \infty} \frac{1}{T} \sum_{\mathcal{A} \in \bmath{\mathcal{A}} \; : \; i \in \mathcal{U}_{\mathcal{A}}} C_i^{\mathcal{A}}(T)}{\displaystyle \lim_{T \longrightarrow \infty} \frac{1}{T} \sum_{\mathcal{A} \in \bmath{\mathcal{A}} \; : \; i \in \mathcal{U}_{\mathcal{A}}} A_i^{\mathcal{A}}(T)} \nonumber \\
&=& \frac{\displaystyle \lim_{T \longrightarrow \infty} \frac{1}{T} \sum_{\mathcal{A} \in \bmath{\mathcal{A}} \; : \; i \in \mathcal{U}_{\mathcal{A}}} B_i^{\mathcal{A}}(T) \times \frac{C_i^{\mathcal{A}}(T)}{B_i^{\mathcal{A}}(T)}}{\displaystyle \lim_{T \longrightarrow \infty} \frac{1}{T} \sum_{\mathcal{A} \in \bmath{\mathcal{A}} \; : \; i \in \mathcal{U}_{\mathcal{A}}} B_i^{\mathcal{A}}(T) \times \frac{A_i^{\mathcal{A}}(T)}{B_i^{\mathcal{A}}(T)}} \nonumber \\ 
&=& \frac{\displaystyle \sum_{\mathcal{A} \in \bmath{\mathcal{A}} \; : \; i \in \mathcal{U}_{\mathcal{A}}} \left(\lim_{T \longrightarrow \infty} \frac{B_i^{\mathcal{A}}(T)}{T}\right) \times \left(\lim_{T \longrightarrow \infty}\frac{C_i^{\mathcal{A}}(T)}{B_i^{\mathcal{A}}(T)}\right)}{\displaystyle \sum_{\mathcal{A} \in \bmath{\mathcal{A}} \; : \; i \in \mathcal{U}_{\mathcal{A}}} \left(\lim_{T \longrightarrow \infty} \frac{B_i^{\mathcal{A}}(T)}{T}\right) \times \left(\lim_{T \longrightarrow \infty} \frac{A_i^{\mathcal{A}}(T)}{B_i^{\mathcal{A}}(T)}\right)} \nonumber \\ 
%&=& \frac{\displaystyle \sum_{\mathcal{A} \in \bmath{\mathcal{A}} \; : \; i \in \mathcal{U}_{\mathcal{A}}} \left(\lim_{T \longrightarrow \infty} \frac{B_i^{\mathcal{A}}(T)}{T}\right) \times \left(\lim_{T \longrightarrow \infty} \frac{A_i^{\mathcal{A}}(T)}{B_i^{\mathcal{A}}(T)}\right) \times \left(\lim_{T \longrightarrow \infty}\frac{C_i^{\mathcal{A}}(T)}{A_i^{\mathcal{A}}(T)}\right)}{\displaystyle \sum_{\mathcal{A} \in \bmath{\mathcal{A}} \; : \; i \in \mathcal{U}_{\mathcal{A}}} \left(\lim_{T \longrightarrow \infty} \frac{B_i^{\mathcal{A}}(T)}{T}\right) \times \left(\lim_{T \longrightarrow \infty} \frac{A_i^{\mathcal{A}}(T)}{B_i^{\mathcal{A}}(T)}\right)} \nonumber \\ 
%&=& \frac{\displaystyle \sum_{\mathcal{A} \in \bmath{\mathcal{A}} \; : \; i \in \mathcal{U}_{\mathcal{A}}} \left(\frac{\pi(\mathcal{A})} \sigma}\right) \times \beta_i \times \left(1 - (1-\beta_i)^{n_i-1} \prod_{j \in \mathcal{N}_i \; : \; j \in \mathcal{U}_{\mathcal{A}}} (1-\beta_j)^{n_j} \right)}{\displaystyle \sum_{\mathcal{A} \in \bmath{\mathcal{A}} \; : \; i \in \mathcal{U}_{\mathcal{A}}} \left(\frac{\pi(\mathcal{A})}{\sigma}\right) \times \beta_i} \nonumber \\ 
%&=& \frac{\displaystyle \sum_{\mathcal{A} \in \bmath{\mathcal{A}} \; : \; i \in \mathcal{U}_{\mathcal{A}}} \pi(\mathcal{A}) \left(1 - (1-\beta_i)^{n_i-1} \prod_{j \in \mathcal{N}_i \; : \; j \in \mathcal{U}_{\mathcal{A}}} (1-\beta_j)^{n_j} \right)}{\displaystyle \sum_{\mathcal{A} \in \bmath{\mathcal{A}} \; : \; i \in \mathcal{U}_{\mathcal{A}}} \pi(\mathcal{A})} \nonumber \\
\end{eqnarray}
%\end{minipage}
%\end{small}
%%\vspace{-5mm}
%%\caption{Derivation of Equation \ref{eqn:gamma_i-multicell}.}
%\end{center}
%\end{figure*}
We can write 
\begin{eqnarray}
\label{eqn:product-of-limits}
\lim_{T \longrightarrow \infty}\frac{C_i^{\mathcal{A}}(T)}{B_i^{\mathcal{A}}(T)} &=& \lim_{T \longrightarrow \infty} \frac{A_i^{\mathcal{A}}(T)}{B_i^{\mathcal{A}}(T)} \times \lim_{T \longrightarrow \infty}\frac{C_i^{\mathcal{A}}(T)}{A_i^{\mathcal{A}}(T)}
\end{eqnarray}
where $\lim_{T \longrightarrow \infty}\frac{C_i^{\mathcal{A}}(T)}{A_i^{\mathcal{A}}(T)}$ can be obtained by 
\begin{equation}
\label{eqn:collision-probability-in-state-A}
\displaystyle \lim_{T \longrightarrow \infty} \frac{C_i^{\mathcal{A}}(T)}{A_i^{\mathcal{A}}(T)} = 1 - (1-\beta_i)^{n_i-1} \prod_{j \in \mathcal{N}_i \; : \; j \in \mathcal{U}_{\mathcal{A}}} (1-\beta_j)^{n_j}. 
\end{equation}

Equation \eqref{eqn:collision-probability-in-state-A} can be explained
as follows. Given that a node in Cell-$i$ attempts a transmission, the
transmission is successful if: (i) the $n_i-1$ other nodes in Cell-$i$
do not attempt transmissions in the same backoff slot, which happens
with probability $(1-\beta_i)^{n_i-1}$, and (ii) for every neighboring
cell, Cell-$j$, that is also in backoff in the state $\mathcal{A}$,
the $n_j$ nodes in Cell-$j$ do not attempt transmissions in the same
slot, which happens with probability $\prod_{j \in \mathcal{N}_i \; :
  \; j \in \mathcal{U}_{\mathcal{A}}}(1-\beta_j)^{n_j}$. Every
transmission attempt by a node in Cell-$i$ in State-$\mathcal{A}$
collides with the complementary probability $p_{coll,\mathcal{A}}$
given by 
\[\quad \quad p_{coll,\mathcal{A}} := 1 - (1-\beta_i)^{n_i-1} \prod_{j \in
  \mathcal{N}_i \; : \; j \in \mathcal{U}_{\mathcal{A}}}
(1-\beta_j)^{n_j}.\] 
Thus, for $T$ sufficiently long, we have 
\[\quad \quad \quad \quad \quad \quad C_i^{\mathcal{A}}(T) \approx p_{coll,\mathcal{A}} A_i^{\mathcal{A}}(T).\]
Equation \eqref{eqn:collision-probability-in-state-A} is obtained by taking the
limit $T \longrightarrow \infty$. 

Combining \eqref{eqn:number-of-backoff-slots}, \eqref{eqn:beta-definition}, \eqref{eqn:gamma_i-multicell-derivation}, \eqref{eqn:product-of-limits}, \eqref{eqn:collision-probability-in-state-A}, we obtain \eqref{eqn:gamma_i-multicell}.

\bibliographystyle{./IEEEtran}
\bibliography{manojPhDbib}

\begin{thebibliography}{10}
\providecommand{\url}[1]{#1}
\csname url@samestyle\endcsname
\providecommand{\newblock}{\relax}
\providecommand{\bibinfo}[2]{#2}
\providecommand{\BIBentrySTDinterwordspacing}{\spaceskip=0pt\relax}
\providecommand{\BIBentryALTinterwordstretchfactor}{4}
\providecommand{\BIBentryALTinterwordspacing}{\spaceskip=\fontdimen2\font plus
\BIBentryALTinterwordstretchfactor\fontdimen3\font minus
  \fontdimen4\font\relax}
\providecommand{\BIBforeignlanguage}[2]{{%
\expandafter\ifx\csname l@#1\endcsname\relax
\typeout{** WARNING: IEEEtran.bst: No hyphenation pattern has been}%
\typeout{** loaded for the language `#1'. Using the pattern for}%
\typeout{** the default language instead.}%
\else
\language=\csname l@#1\endcsname
\fi
#2}}
\providecommand{\BIBdecl}{\relax}
\BIBdecl

\bibitem{wanet.manoj-anuragRAWNET09multicell}
M.~K. Panda and A.~Kumar, ``{M}odeling {M}ulti-{Cell} {IEEE} 802.11 {WLAN}s
  with {A}pplication to {C}hannel {A}ssignment,'' in \emph{RAWNET/WNC3 2009
  Workshop, WiOpt'09}, 2009.

\bibitem{wanet.IEEE802dot11standard2007}
``Wireless {LAN} {M}edium {A}ccess {C}ontrol ({MAC}) and {P}hysical {L}ayer
  ({PHY}) {S}pecifications, {IEEE} {S}td 802.11-2007,'' June 2007.

\bibitem{wanet.bianchi00performance}
G.~Bianchi, ``Performance {A}nalysis of the {IEEE} 802.11 {D}istributed
  {C}oordination {F}unction,'' \emph{IEEE Journal on Selected Areas in
  Communications}, vol.~18, no.~3, pp. 535--547, March 2000.

\bibitem{wanet.kumar_etal07new_insights}
A.~Kumar, E.~Altman, D.~Miorandi, and M.~Goyal, ``New insights from a fixed
  point analysis of single cell {IEEE}~802.11 {WLAN}s,'' \emph{IEEE/ACM
  Transactions on Networking}, vol.~15, no.~3, pp. 588--601, June 2007.

\bibitem{wanet.boorstyn87multihop}
R.~Boorstyn, A.~Kershenbaum, B.~Maglaris, and V.~Sahin, ``{T}hroughput
  {A}nalysis in {M}ultihop {CSMA} {P}acket {R}adio {N}etworks,'' \emph{IEEE
  Transactions on Communications}, vol.~35, no.~3, pp. 267--274, March 1987.

\bibitem{wanet.wang-kar05multihop}
X.~Wang and K.~Kar, ``{T}hroughput {M}odeling and {F}airness {I}ssues in
  {CSMA/CA} {B}ased {A}d {H}oc {N}etworks,'' in \emph{IEEE INFOCOM}, 2005.

\bibitem{wanet.garetto_etal08starvation}
M.~Garetto, T.~Salonidis, and E.~W. Knightly, ``Modeling {P}er-flow
  {T}hroughput and {C}apturing {S}tarvation in {CSMA} {M}ulti-hop {W}ireless
  {N}etworks,'' \emph{IEEE/ACM Transactions on Networking}, vol.~16, no.~4, pp.
  864--877, August 2008.

\bibitem{wanet.kershenbaum-etal87complex}
A.~Kershenbaum, R.~R. Boorstyn, and M.~S. Chen, ``{A}n {A}lgorithm for
  {E}valuation of {T}hroughput in {M}ultihop {P}acket {R}adio {N}etworks with
  {C}omplex {T}opologies,'' \emph{IEEE Journal on Selected Areas in
  Communications}, vol. SAC-5, no.~6, pp. 1003--1012, July 1987.

\bibitem{wanet.durvy09selfOrganization}
M.~Durvy, O.~Dousse, and P.~Thiran, ``{S}elf-{O}rganization {P}roperties of
  {CSMA}/{CA} {S}ystems and {T}heir {C}onsequences on {F}airness,'' \emph{IEEE
  Transactions on Information Theory}, vol.~55, no.~3, March 2009.

\bibitem{wanet.bonald08multicellprocsharing}
T.~Bonald, A.~Ibrahim, and J.~Roberts, ``{T}raffic {C}apacity of {M}ulti-{C}ell
  {WLANs},'' in \emph{ACM SIGMETRICS}, 2008.

\bibitem{wanet.jiang-liew08MobComp-HNEN}
L.~Jiang and S.~C. Liew, ``{I}mproving {T}hroughput and {F}airness by
  {R}educing {E}xposed and {H}idden {N}odes in 802.11 {N}etworks,'' \emph{IEEE
  Transactions on Mobile Computing}, vol.~7, no.~1, pp. 34--49, 2008.

\bibitem{wanet.liew_etal09mobicom-capacity-wireless-networks}
C.~K. Chau, M.~Chen, and S.~C. Liew, ``{C}apacity of {L}arge-scale {CSMA}
  {W}ireless {N}etworks,'' in \emph{MobiCom}, 2009, detailed technical report
  available at {{\tt http://arxiv.org/pdf/0909.3356v4.pdf}}.

\bibitem{wanet.liew_etal09ICCback-of-the-envelope}
S.~C. Liew, C.~Kai, J.~Leung, and B.~Wong, ``{B}ack-of-the-{E}nvelope
  {C}omputation of {T}hroughput {D}istribution in {CSMA} {W}ireless
  {N}etworks,'' in \emph{ICC}, 2009, detailed technical report available at
  {\tt http://arxiv.org/pdf/0712.1854.pdf}.

\bibitem{wanet.nguyen07stochGeom}
H.~Q. Nguyen, F.~Baccelli, and D.~Kofman, ``{A} {S}tochastic {G}eometry
  {A}nalysis of {D}ense {IEEE} 802.11 {N}etworks,'' in \emph{IEEE INFOCOM},
  2007.

\bibitem{wanet.harsha07WiNet}
G.~Kuriakose, S.~Harsha, A.~Kumar, and V.~Sharma, ``Analytical {M}odels for
  {C}apacity {E}stimation of {IEEE} 802.11 {WLAN}s using {DCF} for {I}nternet
  {A}pplication,'' \emph{Wireless Networks (Springer)}, vol.~15, no.~2, pp.
  259--277, February 2009.

\bibitem{wanet.bruno08TCPeqvSatModel}
R.~Bruno, M.~Conti, and E.~Gregori, ``{A}n accurate closed-form formula for the
  throughput of long-lived {TCP} connections in {IEEE} 802.11 {WLANs},''
  \emph{Computer Networks (Elsevier)}, vol.~52, pp. 199--212, 2008.

\bibitem{wanet.litjens_etalITC03integrated_packet_flow}
R.~Litjens, F.~Roijers, J.~L. van~den Berg, R.~J. Boucherie, and M.~Fleuren,
  ``{P}erformance {A}nalysis of {W}ireless {LAN}s: {A}n {I}ntegrated
  {P}acket/{F}low {L}evel {A}pproach,'' in \emph{ITC 18}, Berlin, Germany,
  2003.

\bibitem{wanet.miorandi_etal06http_over_wlans}
D.~Miorandi, A.~A. Kherani, and E.~Altman, ``A {Q}ueueing {M}odel for {HTTP}
  {T}raffic over {IEEE} 802.11 {WLAN}s,'' \emph{Computer Networks (Elsevier)},
  vol.~50, pp. 63--79, 2006.

\bibitem{wanet.bonald_etal04multicell-cellular}
T.~Bonald, S.~Borst, N.~Hegde, and A.~Proutiere, ``{W}ireless {D}ata
  {P}erformance in {M}ulti-{C}ell {S}cenarios,'' in \emph{ACM
  SIGMETRICS/Performance'04}, 2004.

\bibitem{wanet.roy_etal09ToN-PCS}
H.~Ma, R.~Vijayakumar, S.~Roy, and J.~Zhu, ``{O}ptimizing 802.11 {W}ireless
  {M}esh {N}etworks {B}ased on {P}hysical {C}arrier {S}ensing,'' \emph{IEEE/ACM
  Transactions on Networking}, vol.~17, no.~5, pp. 1550--1563, 2009.

\bibitem{queueing.whittle85partial-balance-insensitivity}
P.~Whittle, ``{P}artial {B}alance and {I}nsensitivity,'' \emph{Journal of
  Applied Probability}, vol.~22, pp. 168--176, 1985.

\bibitem{theory.kelly79reversibility}
F.~P. Kelly, \emph{Reversibility and Stochastic Networks}.\hskip 1em plus 0.5em
  minus 0.4em\relax John Wiley, 1979.

\bibitem{wanet.ns2}
S.~McCanne and S.~Floyd, ``The ns {N}etwork {S}imulator (v2.33),'' 2008, {\tt
  http://www.isi.edu/nsnam/ns/}.

\end{thebibliography}
%\vspace*{-0.5em}

\end{document}